%% file: paper.tex
\documentclass[twocolumn,times,tighten]{aastex62}

\usepackage{xspace}
\usepackage{mathptmx}
\usepackage{natbib}
\usepackage{booktabs}
\usepackage{layout}
\usepackage{amsmath}

\newcommand{\six}{4U~1626\ensuremath{-}67\xspace}
\newcommand{\msol}{\ensuremath{M_{\odot}}\xspace}

\newcommand{\suz}{\emph{Suzaku}\xspace}
\newcommand{\nustar}{\emph{NuSTAR}\xspace}

\newcommand{\chandra}{\emph{Chandra}\xspace}

\newcommand{\xmm}{\emph{XMM-Newton}\xspace}
\newcommand{\ciaover}{4.9\xspace}
\newcommand{\caldbver}{4.7.6\xspace}
\newcommand{\isisver}{1.6.2-40\xspace}
\newcommand{\nh}{\ensuremath{N_{\mathrm{H}}}\xspace}
\newcommand{\alfven}{Alfv\'{e}n\xspace}
\newcommand{\Rm}[1]{\uppercase\expandafter{\romannumeral #1\relax}}
\newcommand{\ionrange}[3]{\ion{#1}{#2}--{\small \Rm{#3}}}
\newcommand{\nex}{\ion{Ne}{10}\xspace}
\newcommand{\neix}{\ion{Ne}{9}\xspace}
\newcommand{\nelines}{\ionrange{Ne}{9}{10}\xspace}
\newcommand{\oviii}{\ion{O}{8}\xspace}
\newcommand{\ovii}{\ion{O}{7}\xspace}
\newcommand{\olines}{\ionrange{O}{7}{8}\xspace}
\newcommand{\cvi}{\ion{C}{6}\xspace}
\newcommand{\cv}{\ion{C}{5}\xspace}

\newcommand{\A}{\AA\xspace}
\newcommand{\kms}{km\,s\ensuremath{^{-1}}\xspace}
\newcommand{\gs}{g\,s\ensuremath{^{-1}}\xspace}
\newcommand{\ergs}{erg\,s\ensuremath{^{-1}}\xspace}
\newcommand{\msolyr}{\msol\,yr\ensuremath{^{-1}}\xspace}
\newcommand{\gmc}{\ensuremath{GM/c^{2}}\xspace}
\newcommand{\pflx}{ph\,cm\ensuremath{^{-2}}\,s\ensuremath{^{-1}}\xspace}
\newcommand{\eflx}{erg\,cm\ensuremath{^{-2}}\,s\ensuremath{^{-1}}\xspace}

\begin{document}

\title{\chandra/LETGS Studies of the Collisional Plasma in \six}

\author[0000-0002-1676-6954]{Paul~B.~Hemphill}
\altaffiliation{Current address: Boston Fusion Corp., 70 Westview St Suite 100, Lexington,\\ MA 02421, USA}
\author{Norbert~S.~Schulz}
\author[0000-0002-6492-1293]{Herman~L.~Marshall}
\author[0000-0001-8804-8946]{Deepto~Chakrabarty}
\affiliation{Kavli Institute for Astrophysics and Space Research, Massachusetts Institute of Technology, 77 Massachusetts Ave, Cambridge, MA 02139, USA}

\email{pbh@space.mit.edu}

\begin{abstract}
  We present an analysis of \chandra/LETGS observations of the ultracompact
  X-ray binary (UCXB) \object[4U 1626$-$67]{\six}, continuing our project to analyze the existing
  \chandra gratings data of this interesting source. The extremely low mass,
  hydrogen-depleted donor star provides a unique opportunity to study the
  properties and structure of the metal-rich accreted plasma. There are strong,
  double-peaked emission features of \olines and \nelines, but no other
  identified emission lines are detected. Our spectral fit simultaneously
  models the emission line profiles and the plasma parameters, using a
  two-temperature collisionally-ionized plasma. Based on our line profile
  fitting, we constrain the inclination of the system to 25--60$^{\circ}$ and
  the inner disk radius to $\sim$1500 gravitational radii, in turn constraining
  the donor mass to $\lesssim$0.026\,\msol, while our plasma modeling confirms
  previous reports of high neon abundance in the source, establishing a Ne/O
  ratio in the system of $0.47 \pm 0.04$, while simultaneously estimating a
  very low Fe/O ratio of $0.0042 \pm 0.0008$ and limiting the Mg/O ratio to
  less than 1\% by number. We discuss these results in light of previous work.
 \end{abstract}


\section{Introduction} \label{sec:intro}

\six is an ultracompact X-ray binary (UCXB), featuring a neutron star in a
42-minute orbit around a very low-mass companion, and is unique as the only UCXB
hosting a persistent X-ray pulsar. The source was discovered by \textit{Uhuru}
\citep{giacconi_uhuru_1972}, with 7.68\,s pulsations discovered in SAS-3 data
by \citet{rappaport_discovery_1977}. A faint, blue star was suggested for the
companion by \citet{mcclintock_optical_1977}, confirmed by the observation of
optical pulsations at the X-ray period by \citet{ilovaisky_discovery_1978}. A
highly compact binary orbit with a low-mass companion was first suggested by
\citet{joss_accreting_1978}, based on the lack of significant timing residuals
in the X-ray band. The compactness of the orbit was more firmly established by
\citet{middleditch_4u_1981}, who identified the 42\,min orbital period in the
optical pulsations; this was later confirmed by
\citet{chakrabarty_high-speed_1998}. A cyclotron resonance scattering feature
(CRSF) was discovered at $\sim$37\,keV in \textit{BeppoSAX} data by
\citet{orlandini_bepposax_1998}, implying a magnetic field strength of
$\sim$4.2$\times10^{12}$\,G.

The X-ray pulsar was in a spin-up state with a timescale of $\sim$5000\,yr for
several decades until 1990, after which \textit{CGRO}/BATSE observations
revealed that the source had transitioned to a spin-down state \citep[see][and
references therein]{chakrabarty_torque_1997}. The source's X-ray luminosity had
also been declining steadily since its discovery, which lead
\citet{krauss_high-resolution_2007} to predict that the source would fade into
quiescence in a few years. However, both of these trends were reversed with a
second torque reversal in 2008 \citep{camero-arranz_new_2010}, which was
accompanied by a large increase in X-ray flux \citep{camero-arranz_4u_2012},
bringing the source back to its pre-1990 brightness.

The nature of the donor star in \six is uncertain. Based on the small semimajor axis
\citep[$a\sin i \lesssim 8$\,lt-ms, per][]{shinoda_discovery_1990} and the
requirement that the donor fill its Roche lobe,
\citet{verbunt_evolutionary_1990} and \citet{chakrabarty_high-speed_1998}
proposed that the donor was either a helium-burning star of mass
$\sim$0.6\,\msol, a partially-degenerate, H-poor $\sim$0.08\,\msol star, or a
white dwarf (WD) of mass $\sim$0.02\,\msol. The low-mass white dwarf case is
currently favored, as it imposes less stringent limits on inclination
($\lesssim 33^{\circ}$) and is easier to reconcile with the system's overall
faintness in the optical, blue color, and the lack of detected hydrogen and
helium lines in the system \citep{homer_ultraviolet_2002,werner_vlt_2006}.

The X-ray spectrum from \six is complex and somewhat peculiar, made only more
so by the advent of high-resolution spectroscopy. A complex of emission lines
around 1\,keV were first observed by \citet{angelini_neon_1995} using
\textit{ASCA}; they found anomalously high flux from \nex (the first
hint of enhanced neon abundance in the system), along with emission from
\oviii. High-resolution X-ray spectra from \chandra and \xmm
\citep{schulz_doublepeaked_2001,krauss_high-resolution_2007} resolved the broad
\nex and \oviii features into double-peaked lines, suggesting an accretion disk
origin, as well as finding and characterizing the \neix and \ovii emission. The
only other detected emission feature in \six is a weak iron line complex around
6.4\,keV \citep{camero-arranz_4u_2012}. The 2006 \suz and 2010 \chandra/HETGS
observations report a near-neutral iron fluorescence line at $\sim$6.4\,keV,
while the 2010 \suz observation and 2015 \nustar observations see a 6.7\,keV
feature but no evidence for the neutral line
\citep{camero-arranz_4u_2012,koliopanos_luminosity-dependent_2016,dai_broadband_2017,schulz_collisional_2019}.
It remains unclear why the iron complex appears as different as it does between
these studies --- there were only $\sim$9 months between the 2010 \chandra and
\suz observations. Meanwhile, the optical
\citep{werner_vlt_2006,nelemans_optical_2006} and UV
\citep{homer_ultraviolet_2002} spectra show a complete lack of lines from
hydrogen, helium, and nitrogen, while finding excess emission from oxygen and
some hints of extra carbon in the system. Oddly, \citet{werner_vlt_2006} also
do not detect any neon lines, placing an upper limit on the neon abundance of
$\sim$10\% solar, in conflict with the enhanced neon abundance implied by the
X-ray data.

The low flux of \six in the early 2000s hindered early efforts to carry out
detailed high-resolution studies of the lines. Fortunately, the 2008 torque reversal was
accompanied by a disproportionately large increase in the line flux, with the
lines brightening by a factor of $\sim$8 \citep[compared to $\sim$3 for the
continuum, as shown using \suz observations bracketing the reversal
by][]{camero-arranz_4u_2012}. This has enabled \citet{schulz_collisional_2019}
to carry out a detailed study of the post-torque reversal \chandra/HETGS
observations, and to compare those results with the pre-reversal HETGS data.
Their work settles on a two-temperature, collisionally-ionized plasma in the
disk as the origin of the strong emission lines. If the lines trace the inner
edge of the disk, the pre- and post-torque-reversal \chandra observations imply
that the disk moved inwards, consistent with the transition to spin-up.

Continuing these efforts to comprehensively analyze the high-resolution
spectra of \six, this paper investigates \chandra/LETGS observations of
\six made in 2014, during the source's post-reversal era. Here, we will extend the
work of \citet{schulz_collisional_2019} by implementing a properly
\emph{disk-blurred} plasma model, unlike the simpler red- and blue-shifted
plasma models in the previous paper, allowing us to simultaneously model the
kinematic and plasma parameters of the accretion disk. We present an overview
of the data and the extraction in Section~\ref{sec:obs}. The lightcurves, pulse
period evolution, and pulse profiles are addressed in Section~\ref{sec:lc}. The
bulk of the study is the spectral modeling in Section~\ref{sec:spectra},
particularly the modeling of the prominent emission lines in \six's spectrum in
Section~\ref{sec:diskline}, \ref{sec:aped}, and \ref{sec:aped_rdblur}. We
discuss the results of these fits and their interpretation in
Section~\ref{sec:discussion}, and summarize the results and compare with other
work in Section~\ref{sec:conclusion}.

\section{Observations and Data Reduction} \label{sec:obs}

\chandra has observed \six multiple times over its mission. In this work, we
focus on \chandra ObsIDs 16636, 15765, and 16637, taken in 2014 between July 8
and 13. These observations use the Low-Energy Transmission Grating Spectrometer
\citep[LETGS;][]{brinkman_description_2000}, which uses the Low Energy
Transmission Gratings (LETG) to disperse X-rays onto the High-Resolution Camera
(HRC). These observations are summarized in Table~\ref{tab:obs} and represent
a total exposure of 131.9\,ks. We extracted spectra and lightcurves using the
CIAO \citet{fruscione2006}  v\ciaover, via the Transmission Grating Catalog and Archive software
\citep[TGCat;][]{huenemoerder_tgcat:_2011}\footnote{\url{http://tgcat.mit.edu/}}.
Since the HRC has no energy resolution, all orders are combined in a single
spectrum for each grating arm. We load ancillary response files (ARFs) and
response matrices (RMFs) through eighth order when analyzing these spectra. The
CALDB version was \caldbver, although we also applied an updated degapping
file\footnote{see
\url{https://cxc.harvard.edu/cal/Hrc/Degap/hrcsdegap_centershift.html}} for the HRC, as
otherwise there is a systematic wavelength offset between the positive and
negative-order spectra.  All extractions and analysis were carried out using
version \isisver of the Interactive Spectral Interpretation System
\citep[ISIS;][]{houck_isis:_2000}.  Unless otherwise stated, all uncertainties
are reported at the 90\% level.

\begin{deluxetable}{lrrrr}
  \tabletypesize{\footnotesize}
  \tablewidth{0pt}
  \tablecaption{Summary of \chandra/LETGS observations of \six \label{tab:obs}}
  \tablehead{ \colhead{ObsID} & \multicolumn{2}{c}{Observation start} & \colhead{Exposure} & \colhead{Average countrate} \\[-2ex]
                              & \colhead{Date} & \colhead{MJD}        & \colhead{ks} & }
  \startdata
  16636 & 2014 July 8   & 56846.66 & $23.0$ & $3.21$ \\
  15765 & 2014 July 11  & 56846.05 & $59.4$ & $3.66$ \\
  16637 & 2014 July 13  & 56846.80 & $49.5$ & $3.65$ \\
  \enddata
\end{deluxetable}

\subsection{Lightcurves and Timing Analysis} \label{sec:lc}

\begin{figure*}
  \plotone{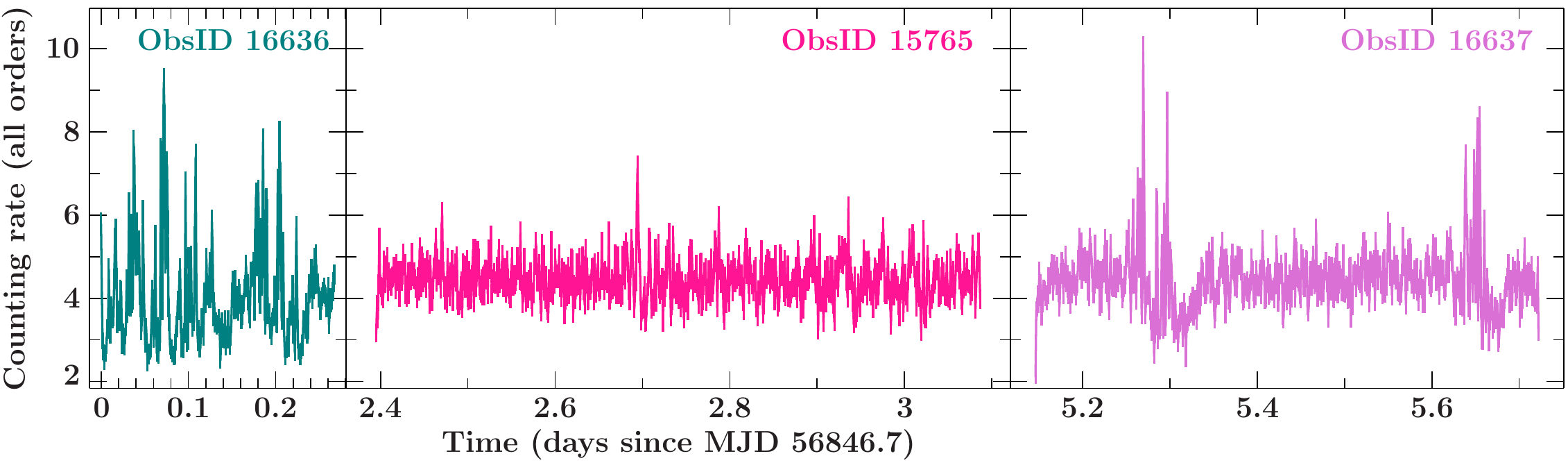}
  \caption{The lightcurves of the three \chandra/LETGS observations of \six, binned
  at $76.7$\,s (ten times the pulse period). The plots are scaled to the
  durations of the observations and are displayed relative to the start time of
  ObsID 16636.}
  \label{fig:lc}
\end{figure*}

We extracted lightcurves for the full LETGS band and for the 0.2--2 and
2--8\,keV bands with 0.5\,s time resolution. The full-band lightcurves of the
three observations are plotted in Figure~\ref{fig:lc}. Between the first and
second observation (ObsIDs 16636 and 15765, respectively), the flux increased
by $\sim$15\%. Variability changes are also clearly visible in the lightcurves,
with the first exhibiting considerably more high-amplitude short-term
variability than the later two, and two (or three) large flares in the last
observation.

\begin{figure}
  \plotone{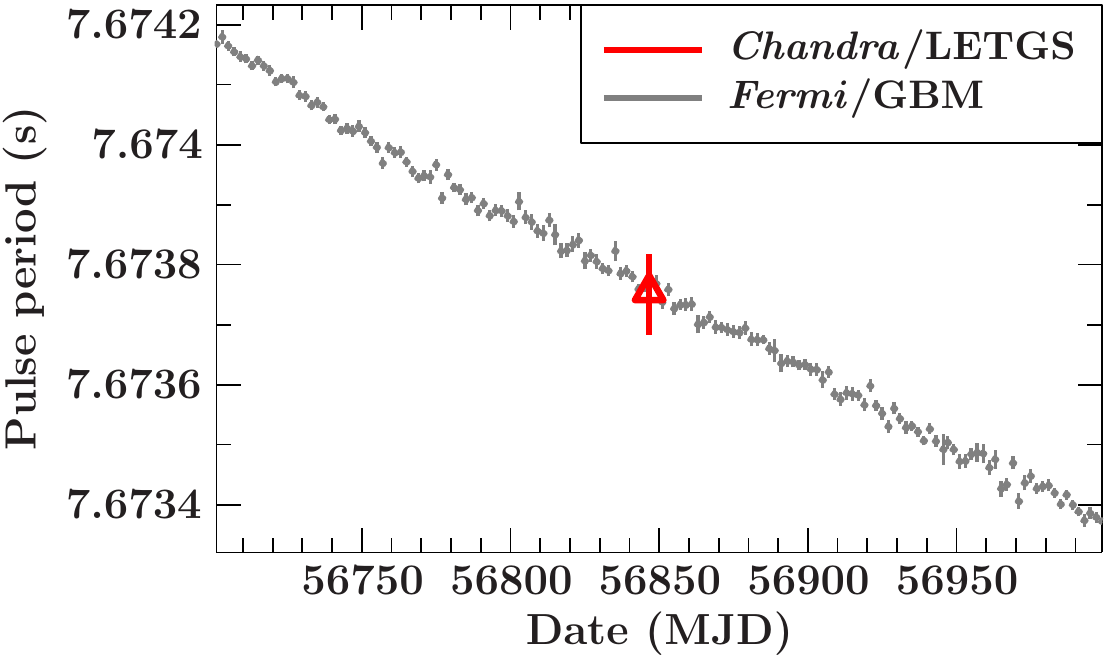}
  \caption{The pulse period determined from the \chandra/LETGS, overplotted with
  the pulse periods measured by the \textit{Fermi}/GBM.}
  \label{fig:pulseperiod}
\end{figure}

After barycentering the lightcurve, a first estimate of the pulse period was
determined via epoch folding \citep{leahy_searches_1983}. A pulse period
determined from epoch folding can be refined by comparing individual pulses at
this period to a reference pulse profile, typically of the full observation.
The phase shifts between the individual pulses and reference profile, $\delta
\varphi$, are related to a correction to the assumed pulse frequency, $\delta
\nu$, and the rate of change of the pulse period, $\dot{\nu}$, via
\begin{equation}
  \delta\varphi\left(t\right) = \varphi_{0} + \delta\nu\left(t-t_{0}\right) + \frac{1}{2}\dot{\nu}\left(t-t_{0}\right)^2,
  \label{eqn:phaseshift}
\end{equation}
where $t_{0}$ is some reference time and $\varphi_{0}$ is the phase offset at
$t = t_{0}$. This allows one to correct the original estimate for the pulse
frequency, and the procedure is repeated until it converges (this typically
only takes a few iterations). The LETGS lightcurves do not have enough counts to
do this on a pulse-by-pulse basis, so we produced average pulse profiles by
adding up every 200 pulses based on the period derived
from epoch folding, and found phase shifts by cross-correlating with the
average pulse profile from the entire LETGS lightcurve. The phase shifts are
nearly linear and the procedure outlined above converges after only a single
iteration. The final pulse period we obtain is $7.67375(7)$\,s, consistent with
the pulse period obtained by \textit{Fermi}/GBM\footnote{See
\url{https://gammaray.nsstc.nasa.gov/gbm/science/pulsars/lightcurves/4u1626.html}
and \citet{finger_long-term_2009}} during this time period
\citep{camero-arranz_new_2010}, as shown in Figure~\ref{fig:pulseperiod}.  The
short time spanned by the LETGS observations means $\dot{\nu}$ cannot be
meaningfully constrained, but our results are consistent with the decline of
$\sim$0.001\,s\,yr$^{-1}$ from the \textit{Fermi} measurements.

\begin{figure}
  \plotone{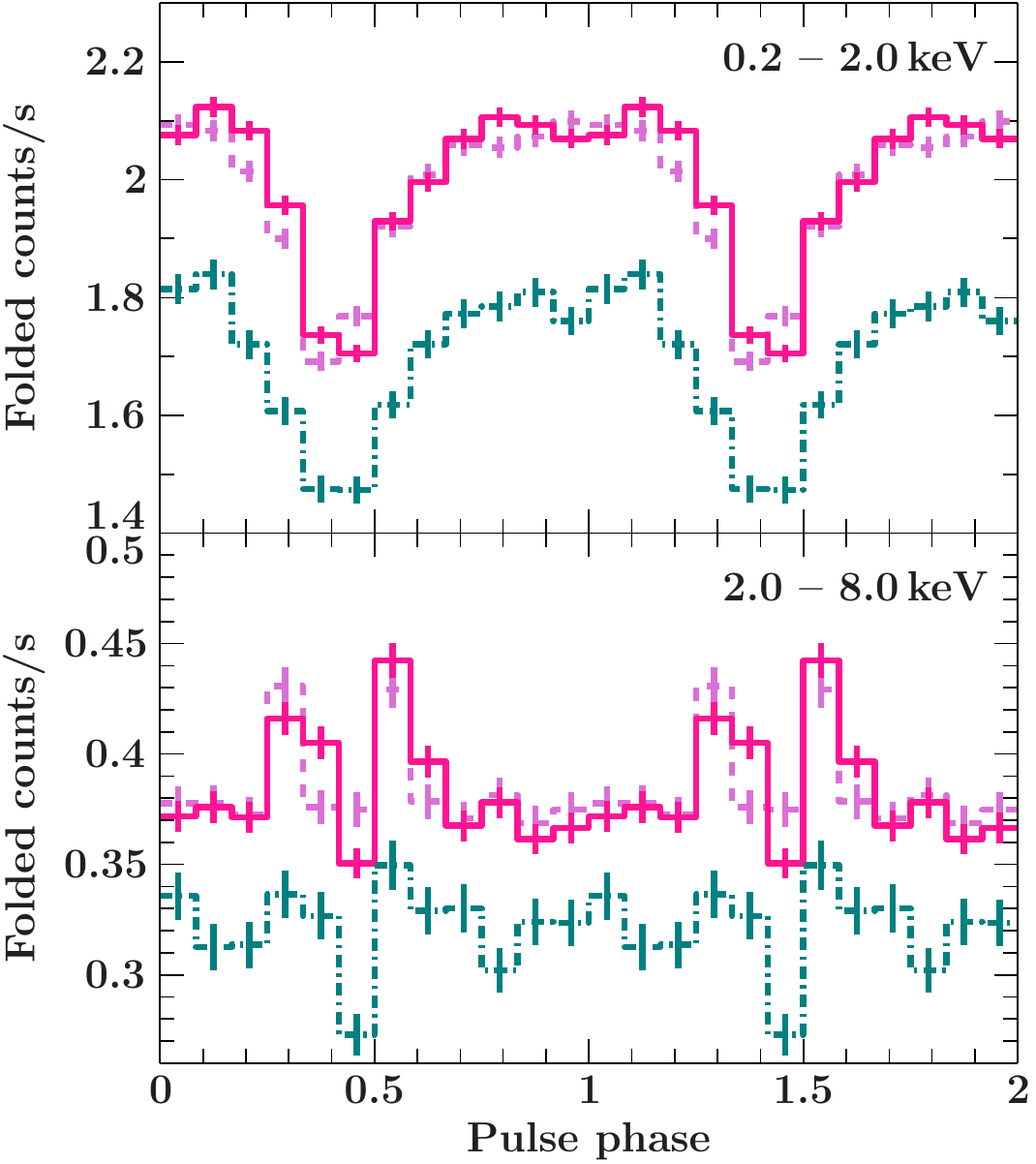}
  \caption{The pulse profiles for \six in the 0.2--2 (top) and 2--8\,keV
  (bottom) bands. ObsID 16626 is the teal dash-dotted line, while 15765 and
  16637 are the solid pink and dashed violet lines, respectively. The dramatic
  change in shape, from a broad single peak to a pair of close, narrow peaks is
  characteristic of this source. The count rate is low in the 2--8\,keV profile
  mainly due to the LETGS' low effective area at higher energies.}
  \label{fig:profs}
\end{figure}

Folding the observations' lightcurves on the measured pulse period obtains the
pulse profiles displayed in Figure~\ref{fig:profs}. The pulse profile changes
from a single broad peak at low energies to a pair of narrow peaks close
together in phase above 2\,keV.  This may be indicative of a switch from a fan
beam at low energies to a pencil beam at higher energies
\citep{pottschmidt_applying_2018}. These profiles are qualitatively very
similar to those presented in \citet{beri_pulse-phase_2015}.

\section{Spectral analysis} \label{sec:spectra}

\begin{figure}
	\plotone{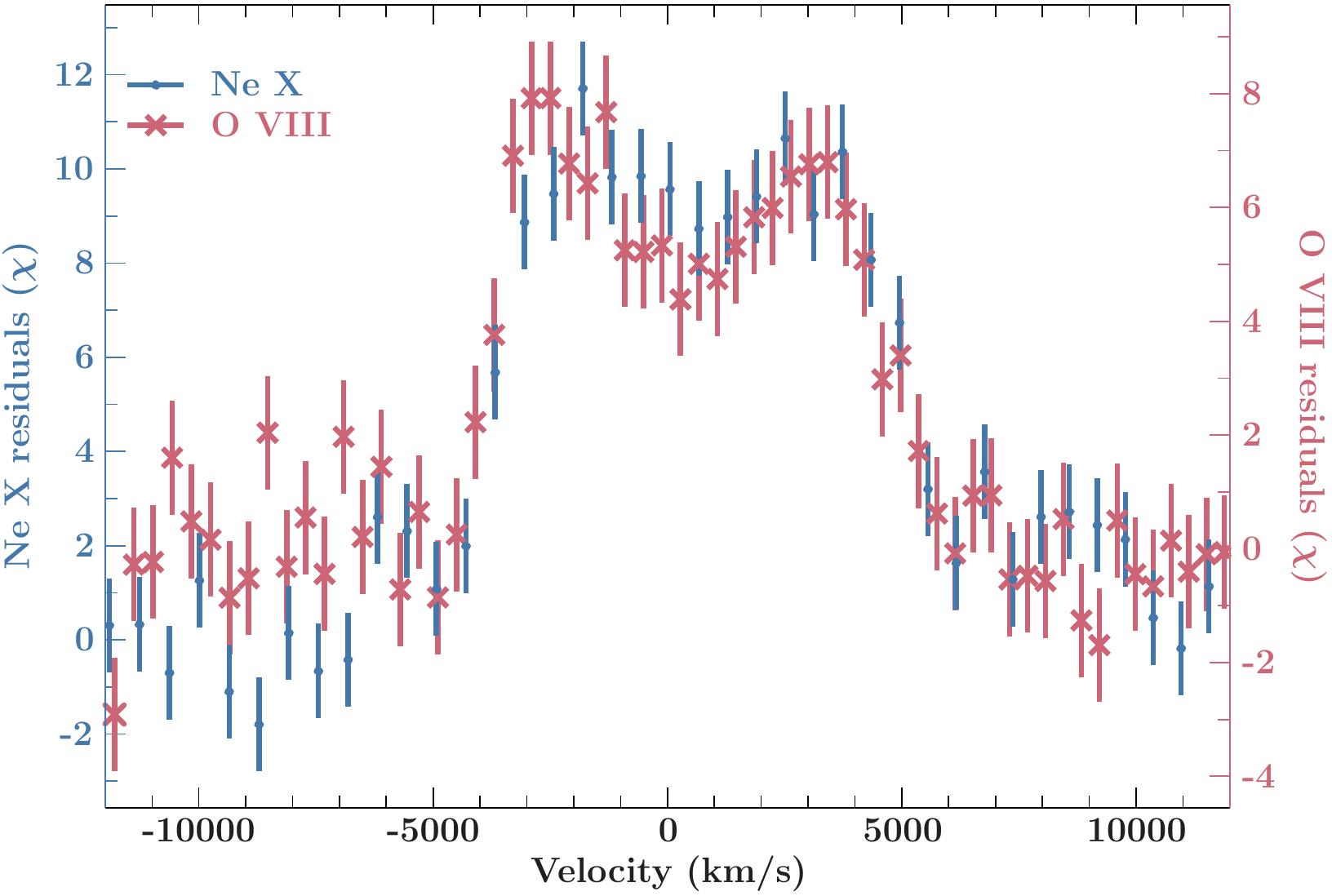}
  \caption{The first-order \nex (blue points) and \oviii (red crosses) line profiles as seen
    in ObsIDs 15765 and 16637. We plot here the residuals after both grating arms have
    been combined and the continuum has been subtracted off, placed on a common
    velocity axis relative to the literature values
    \citep{erickson_energy_1977} for the K$\alpha$ transition for H-like Ne and
    O. The double-peaked nature of the lines is clearly visible, and both
    features show very similar profiles and velocity widths. Note the different
    scales for the lines --- the \nex profile follows the left y-axis scale,
    and the \oviii line follows the right y-axis. Also note that the LETGS
    spectra also have the higher-order counterparts to these lines (not plotted),
    which have accordingly higher spectral resolution.}
    \label{fig:lineprofs}
\end{figure}

The LETGS spectra show unambiguous emission features around $12$, $13$, $19$,
and $21$\,\A, which we identify as emission from \nex, \neix,
\oviii, and \ovii, respectively. Their higher-order counterparts\footnote{Due
to the lack of order-sorting mentioned in Section~\ref{sec:obs}, spectral features
reappear in the spectrum at integer multiples of their wavelength; this is
accounted for by assigning to each spectrum multiple RMFs and ARFs, one for each order.}
are also visible, with the most obvious higher-order feature being the third-order \nex line
appearing at $\sim$36.5\,\A. The features, especially the hydrogen-like lines,
have clear double-peaked profiles and similar widths. Fitting the hydrogen-like
lines with a pair of Gaussians in the manner of
\citet{krauss_high-resolution_2007}, tying the velocity shifts of the red and
blue peaks together, the \nex peaks are shifted relative to their rest
wavelength by $2370 \pm 90$\,\kms, and the \oviii line peaks are shifted by
$2360 \pm 140$\,\kms. These are moderately wider separations than were measured
by \citet{krauss_high-resolution_2007} and consistent with \citet{schulz_collisional_2019}'s conclusion that the line-producing
region has moved inwards since the 2008 torque reversal. Figure~\ref{fig:lineprofs} displays the \nex and \oviii
profiles on a common velocity axis. Interestingly, despite the presence of carbon
in the UV and optical spectra \citep{homer_ultraviolet_2002,werner_vlt_2006},
no emission from \cvi (33.74\,\A) or \cv (40.73--41.472\,\A) is visible. The
apparent lack of carbon may be due to high temperatures in the disk --- this is
discussed in more detail in Section~\ref{sec:discuss_abund}. The only other
immediately apparent feature is an emission feature at $\sim$17\,\A, which we
model in this work with a Gaussian. This feature's identity is not immediately
apparent --- see Section~\ref{sec:17A_line} for a more in-depth discussion.

In the following spectral analysis, unless otherwise noted, we bin the LETGS
spectra to a minimum of 2 channels per bin and a minimum signal-to-noise of 8,
only using data between 2 and 60\,\A. We use $\chi^2$ statistics when fitting,
as with this binning the spectra generally have significantly more than 20
counts per bin.

\subsection{Spectral continuum} \label{sec:continuum}

To establish a spectral model for the continuum, we fitted the combined spectra
from both grating arms, excising data within 0.5\,\A of the \nelines and
\olines lines (and their higher-order counterparts). The continuum model chosen
is the same as has been used in the past for this source: an absorbed $\Gamma
\sim 1$ power law plus a $\sim$0.5\,keV black body.  For broad-band absorption, we use
\texttt{TBabs}, an updated version\footnote{See \citet{wilms_dont_2010} and
\url{pulsar.sternwarte.uni-erlangen.de/wilms/research/tbabs/}} of the
\texttt{tbabs} absorption model of \citet{wilms_absorption_2000}, from which we
also take our ISM abundances. The cross-sections are taken from
\citet{verner_atomic_1996}. The best-fit parameters for each ObsID can be found
in Table~\ref{sec:continuum}.

While there are obvious changes in flux and spectral shape between these three
ObsIDs (most obviously between the first and second observations), these
differences are not straightforwardly reflected in the best-fit spectral
parameters --- all parameters are consistent with each other within their 90\%
uncertainties. We thus carried out a set of Markov Chain Monte Carlo (MCMC) runs to
obtain the approximate probability distribution of the parameter space, using an
ISIS implementation of the \texttt{emcee} algorithm
\citep[produced by M. Nowak, based on][]{foreman-mackey_emcee:_2013}. Each
dataset was run separately with 100 walkers and a chain length of 5000. The
walkers generally settled to their most-probable values within $\sim$500 steps,
so in our analysis we discard the first 10\% of each chain.

\begin{figure*}
  \plotone{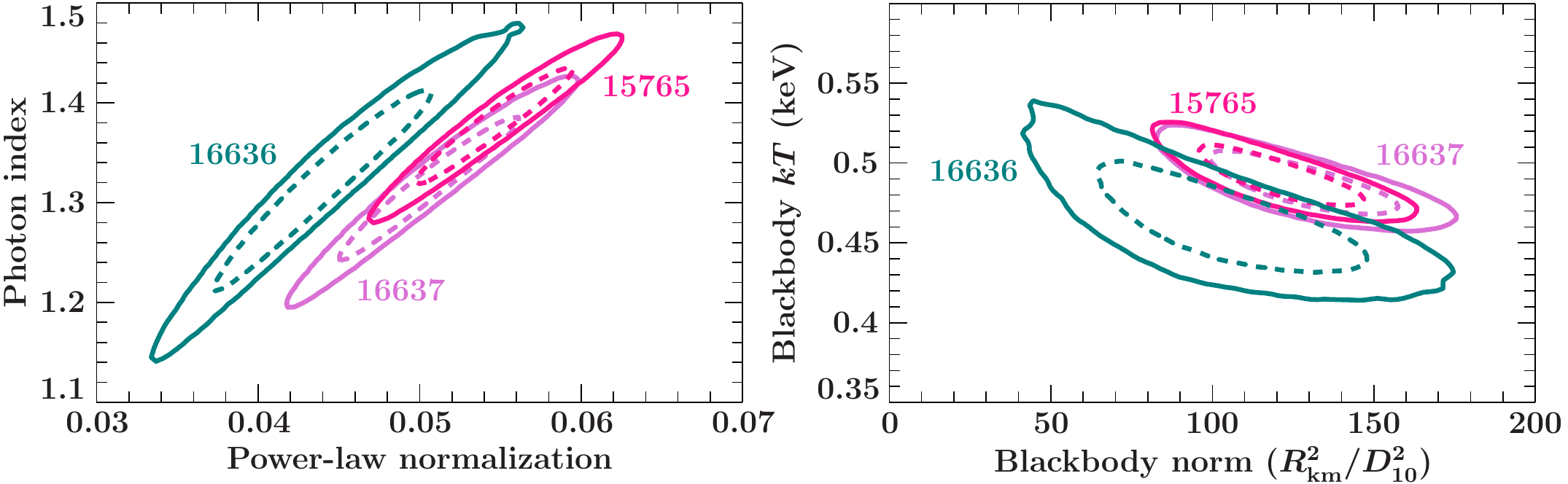}
  \caption{The MCMC-derived 2-dimensional contours for the power law
    normalization vs. power law index (left) and the blackbody temperature vs.
    normalization (right). The contours represent the 68\% and 95\% confidence
    limits. The power law normalizations are inconsistent between the first
    observation (ObsID 16636, teal) and the final two (ObsIDs 15765 and 16637,
    magenta and pink, respectively), while there is a sizable overlap in the
    blackbody parameters.}
  \label{fig:mcmc_contours}
\end{figure*}

In Figure~\ref{fig:mcmc_contours}, we plot the two-dimensional 95\% probability
contours for the power law and blackbody parameters from the MCMC runs. Strong
degeneracies between these parameters result in the large single-parameter
uncertainties and formal consistency between the datasets that were seen in the
spectral fits. However, the MCMC results also reveal a clear inconsistency in the
power law normalization between the first and the later observations. Thus, a
change in the power law normalization (perhaps due to a change in the accretion
rate) appears to be the driver of the change in flux between the observations.
Additionally, despite the overlap in the contours, it is interesting to note that
the positive correlation between the blackbody temperature and area between the
observations runs counter to the long-term inverse correlation seen in the
post-torque-reversal HETGS data \citep{schulz_collisional_2019}.

We also find a significantly higher \nh compared to previous studies
\citep[$\sim 1.7\times 10^{21}$\,cm$^{-2}$, compared to the $\sim 1.4\times
10^{21}$\,cm$^{-1}$ reported by][]{krauss_high-resolution_2007}), and a
moderately softer photon index \citep[$\sim$1.4, compared to $\sim$1.2\
in][]{schulz_collisional_2019}. However, our MCMC runs find strong degeneracy
between \nh and $\Gamma$, and the 95\% confidence contours are consistent with
the values of \nh and $\Gamma$ found by \citet{schulz_collisional_2019} using HETGS
data. Further fits with \nh fixed to $1.4 \times 10^{21}$\,cm$^{-2}$ bear this
out, finding values more consistent with previous studies.

The absorption edges of the main elements in question in this study (C, O, Ne,
Mg, and Fe; see Section~\ref{sec:discuss_donor}) do not appear to be overly
strong in the LETGS spectra. Fitting with a variable-abundance absorption model (\texttt{TBvarabs}) and allowing these
abundances to vary, the oxygen, carbon, and magnesium abundances derived from
the line-of-sight absorption are consistent with their ISM values, although
only oxygen is well-constrained. For carbon the poor constraint may be related
to the fact that the carbon K edge at 43.6\,\A is coincident with the
\ion{O}{7} line in the second order. Neon and iron are moderately
underabundant, with 90\% upper limits of $0.55$ and $0.75$ relative to the ISM
values.

\begin{deluxetable*}{lrrr}
  \tabletypesize{\footnotesize}
  \tablewidth{0pt}
  \tablecaption{Continuum spectral parameters for \six, excluding spectral lines. The source distance is assumed to be 5 kpc (\citet{schulz_collisional_2019})  \label{tab:continuum}}
  \tablehead{& \colhead{16636}&\colhead{15765}&\colhead{16637}}
  \startdata
    $N_{\rm H}$ ($\times 10^{22}$\,cm$^{-2}$) & $0.173\pm0.018$ & $0.191\pm0.020$ & $0.177\pm0.018$ \\ 
    $A_{\rm PL}$ ($\times 10^{-2}$\,\pflx) & $4.3\pm0.5$ & $5.5\pm0.6$ & $5.1\pm0.6$ \\ 
    $\Gamma$ & $1.31\pm0.14$ & $1.38\pm0.14$ & $1.32\pm0.14$ \\ 
    $A_{\rm BB}$ ($R_{\rm km}^2/D_{10}^2$) & $114\pm12$ & $118\pm12$ & $123\pm13$ \\ 
    $kT_{\rm BB}$ (keV) & $0.45\pm0.05$ & $0.49\pm0.05$ & $0.48\pm0.05$ \\ 
    $\lambda_{\rm gauss}$ (\A) & $16.8\pm1.7$ & $16.9\pm1.7$ & $16.9\pm1.7$ \\ 
    $\sigma_{\rm gauss}$ (\A) & $0.215\pm0.022$ & $0.210\pm0.021$ & $0.256\pm0.026$ \\ 
    $A_{\rm gauss}$ ($\times 10^{-4}$\,\pflx) & $5.9\pm0.6$ & $6.0\pm0.7$ & $7.3\pm0.8$ \\ 
    $F_{0.5\text{--}10{\rm keV}}$ ($\times 10^{-10}$\,\eflx) & 
    $4.63$ & $5.47$ & $5.52$ \\
    $L_{0.5\text{--}10{\rm keV}}$ ($\times
    10^{36}$ \ergs) &
    $1.38$ & $1.64$ & $1.65$  \\
  \enddata
\end{deluxetable*}

\subsection{Diskline modeling} \label{sec:diskline}

Building on the work in \citet{schulz_collisional_2019}, we present here a
detailed analysis of of the double-peaked emission lines. As in \citeauthor{schulz_collisional_2019}, we
model the \nelines and \olines lines with the \texttt{diskline} model
\citep{fabian_x-ray_1989}, fixing the emissivity power-law index $q$ to $-3.6$, the same value used for the HETGS data. This large
negative value for $q$ (i.e., line emission falling off quickly with radius) is
justified by the clear trough between the line peaks in both the \nex and
\oviii profiles: if there were significant emission from the outer disk (i.e.,
$q$ closer to zero), it would fill in the space between the peaks.

We obtained a first set of fits using only the \nex and \oviii
features, as they are bright and, unlike the He-like triplets, they are not
blended with other features. We performed simultaneous local fits to the Ne and
O features, using the best-fit continuum from Section~\ref{sec:continuum} and
noticing only the 11--13\,\A and 18--20\,\A regions of the spectra. We
initially fit with the inner and outer disk radii ($R_{\rm in}$ and $R_{\rm
out}$) free to vary between the lines and between observations, while
the inclination was tied between all components. The inner disk radius was
consistent between all ObsIDs and between the \nex and \oviii
lines, at roughly 1000--1500\,\gmc, which for a $1.4$\,\msol neutron star
translates to 2--3$\times 10^{8}$\,cm. The outer radii were more poorly constrained,
which is unsurprising given the steep slope in the emissivity that we assume,
but were typically around $10^{4}$\,\gmc ($2\times10^{9}$\,cm), with
lower bounds of a few $\times 10^{3}$\,\gmc. The line wavelengths are
consistent with their rest wavelengths \citep[from][via
AtomDB]{erickson_energy_1977}, and the oxygen-to-neon flux ratios were
$\sim$1.1\ in all observations. The full set of best-fit parameters can be
found in Table~\ref{tab:diskline_hlike}.

\input{diskline_hlike_local}

\begin{figure*}
  \centering
  % \plottwo doesn't work because it sets the figures based on width, not
  % height, and for some reason LaTeX thinks these two PDFs are different
  % widths, at least as far as plottwo is concerned, meaning they get rescaled
  % differently and don't line up.
  %\plottwo{ne_triplet}{o_triplet}
  \includegraphics[height=8.0cm]{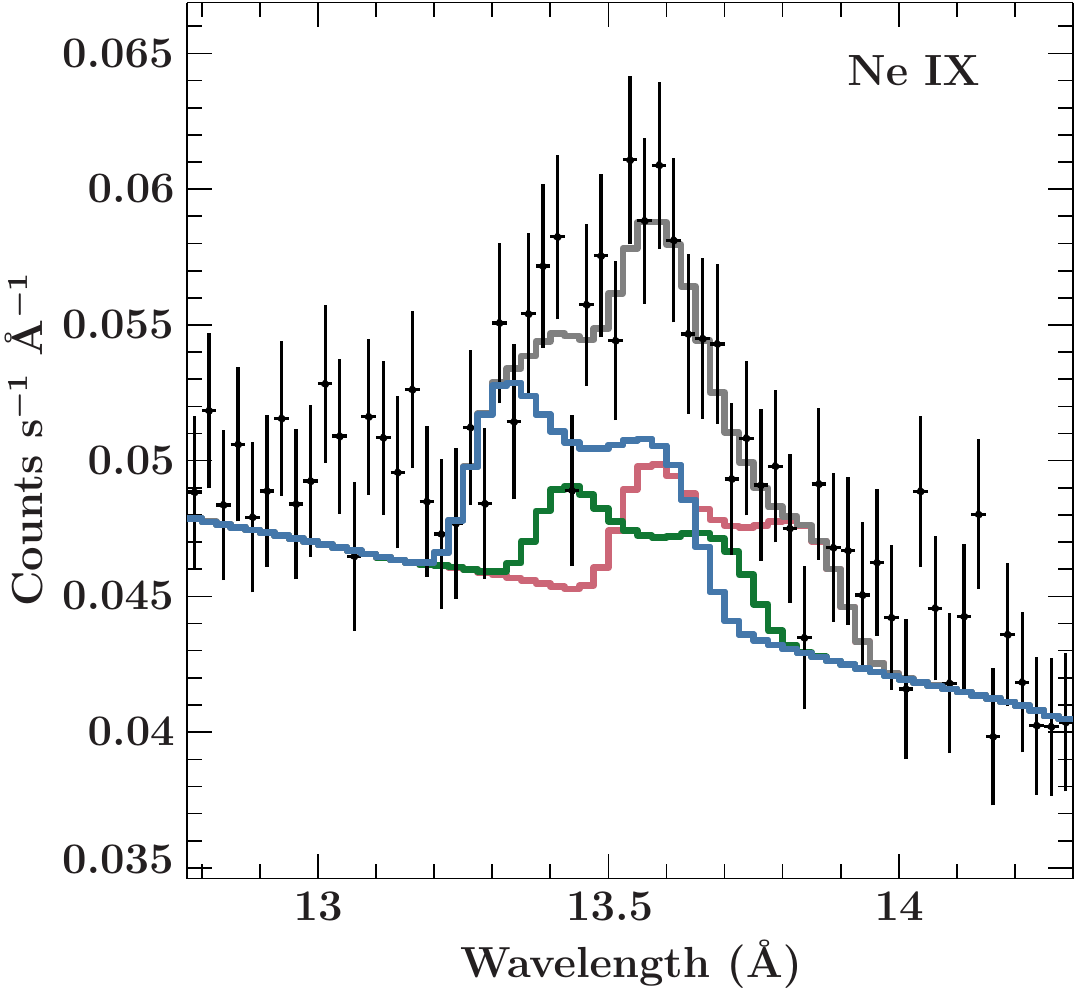}
  \includegraphics[height=8.0cm]{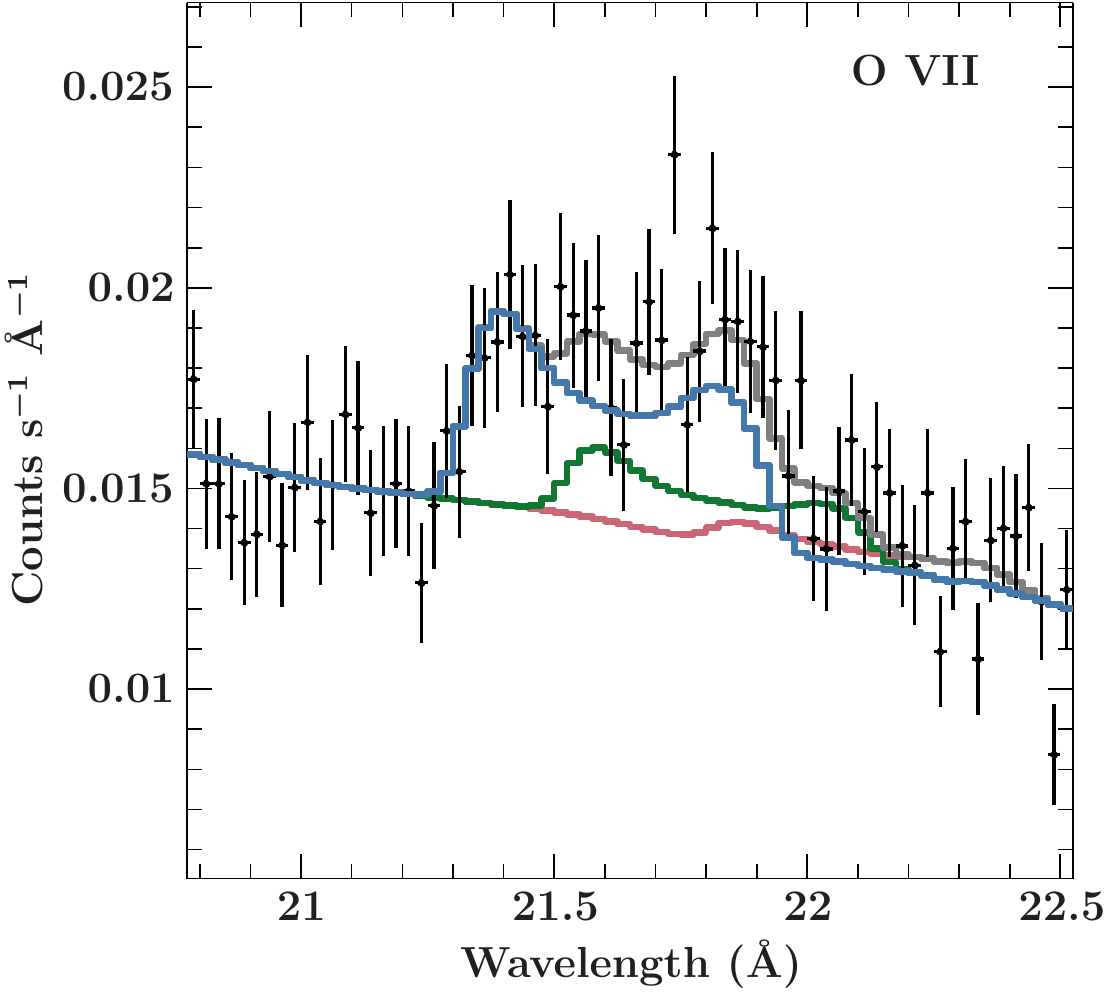}
  \caption{The \neix (left) and \ovii (right) triplets, decomposed into, from
  left to right within each plot, the $r$ (blue), $i$ (green), and $f$ (red)
  transitions. The gray line shows the total model. For clarity, we show all
  ObsIDs combined here. The resonance lines are strong in both profiles,
  dominating the flux of the oxygen triplet, supporting a high-temperature
  collisional plasma as the nature of the underlying medium.}
  \label{fig:triplets}
\end{figure*}

Once the disk parameters were established, we fixed their values and expanded
the wavelength range and model to include the helium-like triplets. Since there
was no detectable shift in the line wavelengths for the H-like features, we
fixed the wavelengths of the \neix and \oviii lines to their values
in AtomDB \citep{Foster2012}, and re-fit to obtain the
normalizations of the He-like triplets. The resulting fits are summarized in
Table~\ref{tab:diskline_helike} and plotted in Figure~\ref{fig:triplets}.

The \texttt{diskline} fits indicate a strong resonance feature compared to the
forbidden and intercombination transitions. This is interesting, as visually
the triplets, especially \neix, appear to be dominated by the intercombination
transition \citep[as reported
by][]{schulz_doublepeaked_2001,krauss_high-resolution_2007,schulz_collisional_2019}. In fact, the velocity
splitting of the lines by the disk is coincidentally similar to the separation between the
lines in the triplets, and as a result the apparent strong ``intercombination'' feature
seen in the \neix triplet is actually the sum of the red wing of the $r$
line and the blue wing of the $f$ line. The $i$ and $f$ transitions are weak
and typically more poorly constrained than the $r$ transitions, especially in
the oxygen triplet.

To estimate confidence intervals for the $R = f/i$ and $G = (f+i)/r$ ratios, we
carried out a set of MCMC runs, with 100 walkers and a chain length of 2000,
(this typically converged to the best-fit parameters $\sim$1000 steps, so we
discarded the first 50\% of the chains). During this process we adjusted the
hard limits on the line amplitudes to allow for negative fluxes. The resulting
posterior distributions of line normalizations for the $r$, $i$, and $f$
transitions for neon and oxygen were convolved to determine the distributions
for $R$ and $G$. We include the maximum \textit{a posteriori} values and 90\%
confidence intervals for $R$ and $G$ in Table~\ref{tab:diskline_helike}. As can
be seen clearly from the confidence intervals, the distributions are highly
skewed, typically with long tails at the high end, due to the skewed distributions of the
line fluxes themselves. The poor detection of the $i$ and $f$ lines in some of
the observations also leads to negative $R$-ratios in some cases, due to the
line fluxes being consistent with zero or negative. Nonetheless, the results
prefer relatively low values (less than a few) for $R$ and $G$. The $R$-ratio
is too poorly determined to offer a meaningful constraint on density
\citep[and, in any case, is likely contaminated by UV depopulation from the
disk, per][]{schulz_collisional_2019}, but the moderately low $G$ value found
in most of the observations favors the hot, collisionally-ionized plasma
inferred by \citet{schulz_collisional_2019} as the origin for the emission.

\input{diskline_helike_local}

\subsection{Plasma modeling} \label{sec:aped}

\begin{figure*}
  \plotone{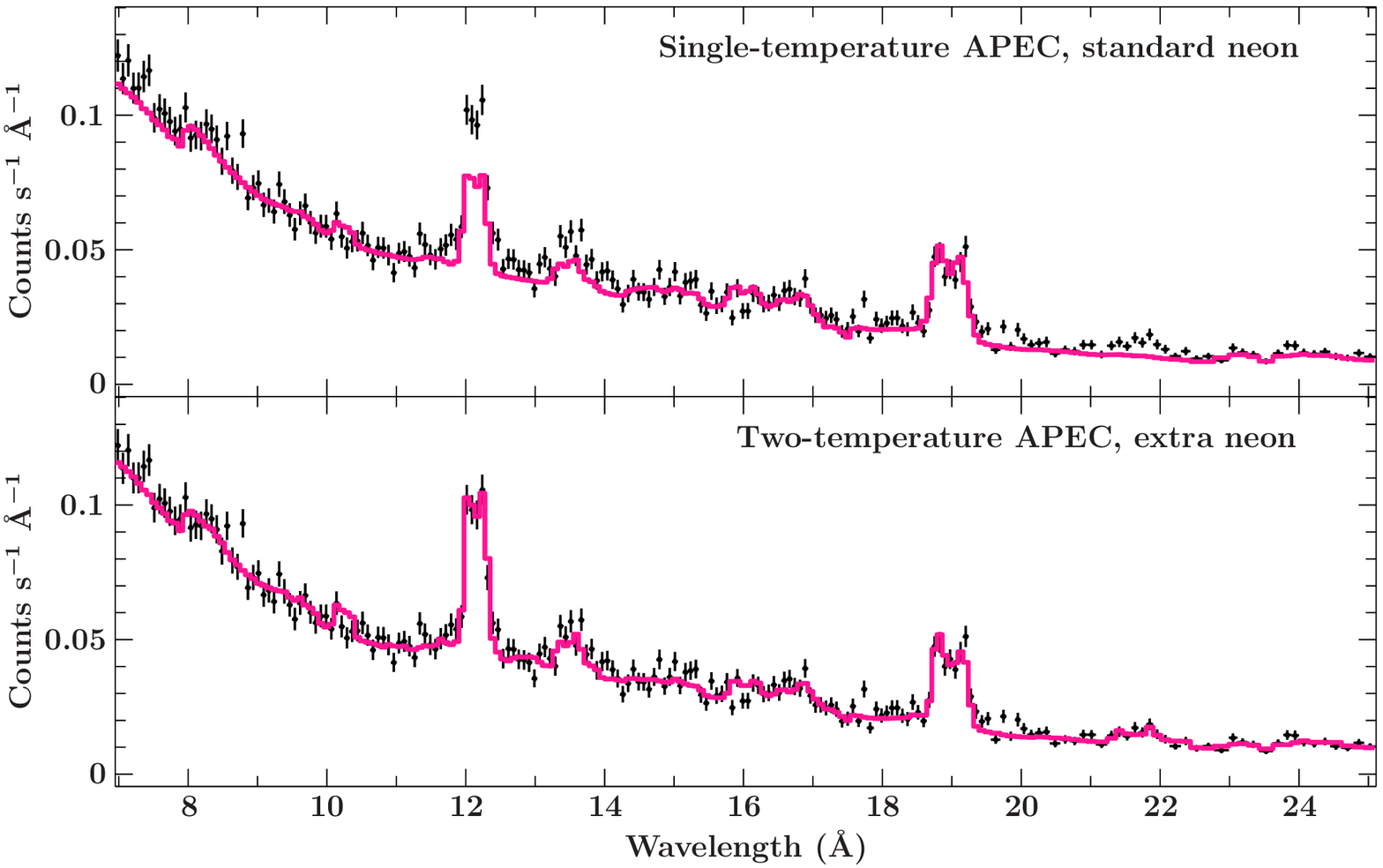}
  \caption{From top to bottom, a two-temperature photoionized plasma with
    enhanced neon, a single-temperature APEC with standard neon abundance, and
    a two-temperature APEC with higher neon abundance. Both panels show ObsIDs
    15765 and 16637 combined. Enhanced neon abundance is clearly needed to
    produce the observed Ne-to-O ratio. The photoionized plasma clearly fails
    to fit the spectrum, with a strong radiative recombination feature from Ne
    around 9\,\A and failing to reproduce the He-like triplets. The
    single-temperature APEC also fails to reproduce the helium-like oxygen triplet
    around 22\,\A.}
  \label{fig:compare_models}
\end{figure*}

Continuing the work of \citet{schulz_collisional_2019}, we constructed a set
of APEC plasma models (functionally equivalent to XSPEC's \texttt{vvapec}
models). As discussed in \citet{schulz_collisional_2019}, the true nature of
the plasma in \six's accretion disk is likely much more complicated than APEC
can provide on its own --- this model is only an approximation. However, we
nonetheless find that we achieve a remarkably good reproduction of the observed
spectra using this approach.

While the baseline abundances used by the APEC models are the solar abundances of
\citet{anders_abundances_1989}, we vary these quite drastically to obtain a
good reproduction of \six's spectrum. First, as in
\citet{schulz_collisional_2019}, we set the abundances of all elements to zero,
other than those of carbon, oxygen, neon, and iron, as only those elements have
detected features in the optical, UV, or X-ray spectra
\citep{homer_ultraviolet_2002,werner_vlt_2006,krauss_high-resolution_2007,camero-arranz_4u_2012},
and the expected donor star is likely depleted in other elements.
Additionally, the conventional definition of ``abundance'' (i.e., measured
relative to the hydrogen number density) is inapplicable to the
hydrogen-depleted plasma we see in \six, so we instead measure abundances
relative to oxygen by fixing the oxygen abundance to unity (see
Section~\ref{sec:discuss_abund} for how to interpret the resulting
measurements).

Based on the temperature inferred from the $G$-ratio above, we initially tried
a 10\,MK plasma with red- and blueshifted components. However, this did not fit
the data adequately. As can be seen in Figure~\ref{fig:compare_models}, a
single-temperature plasma with a neon abundance set to unity (i.e., a solar
Ne/O ratio) fails to fit the relative heights of the neon and oxygen lines,
producing too little flux from neon, and it fails to reproduce the helium-like
triplets from either element. Allowing the neon abundance to vary, we found
that a factor of $\sim$3.5 overabundance of neon successfully reproduced the
high \nex flux and improved the fit to the \neix triplet with a , but the \ovii
triplet could only be fully reproduced by introducing a second,
lower-temperature component with $T_{\rm cool} \sim 2$\,MK.

At these high temperatures, the carbon lines are intrinsically weak (due to the
carbon being almost fully ionized --- see Section~\ref{sec:discuss_abund}), and
as a result we could not obtain meaningful constraints on the carbon abundance.
However, we do find a very low but nonzero iron abundance of $0.025$, mostly
constrained by the lack of detected L-shell iron lines in the 15--20\,\A range
--- the only feature in this range is the unidentified 17\,\A feature, which
this plasma model cannot reproduce without also producing considerably
higher-flux iron lines in the 15--16\,\A region, which are inconsistent with
the data. The LETGS' sensitivity below a few \A is too low to detect the iron K
line, but the upper limits for a line in that region are consistent with the
line seen by \nustar \citep{dai_broadband_2017,iwakiri_spectral_2019}.

The inferred turbulent velocities of the two components were $\sim$2200 and
$\sim$1500\,\kms, similar to what was found in the HETGS data; however, it
should be noted that this does not take into account any blurring due to the
disk. The red- and blueshifts on the 10\,MK and 2\,MK components were
consistent within errors at approximately $\pm 0.008$, suggesting that the two
temperature components exist at roughly the same orbital distance from the
neutron star. This is reassuring, given our assumptions made when modeling the
helium-like lines with \texttt{diskline} above.

\subsection{Combined APEC-\texttt{diskline} Modeling} \label{sec:aped_rdblur}

We can improve on our somewhat crude red- and blueshifted APEC model by using
\texttt{diskline} profiles for the lines in the APEC model. We accomplish this
by convolving the two-temperature APEC with an \texttt{rdblur} model component, which
relativistically blurs an arbitrary input spectrum based on the same code as the
\texttt{diskline} model. Since our \texttt{diskline} modeling and our initial
APEC fits suggest that both temperature components come from approximately the
same orbital distance, we use a single \texttt{rdblur} component for both
temperatures. As in Section~\ref{sec:diskline}, we fix $q$ to $-3.6$, but we
leave the remaining disk parameters free to vary. Our final best-fit model for \six is thus
\texttt{TBabs(pow+bb+rdblur(apec2)+gauss)}, where \texttt{pow} is the power
law, \texttt{bb} is the blackbody, \texttt{gauss} is the Gaussian at 16.9\,\A,
and \texttt{apec2} is a two-temperature APEC model with variable abundances. We allow the power law and
blackbody components to vary between the observations, but tie the plasma and
disk parameters for all ObsIDs together, as our \texttt{diskline} fits found
consistent parameters and line ratios between all observations. As before, we
set the abundances of most elements in the APEC plasma to zero --- although now
in addition to carbon, oxygen, neon, and iron, we also include magnesium, as
this would be the next-most abundant element in an oxygen-neon white dwarf,
which is one of the possible donor stars in this system.

\begin{figure*}
  \plotone{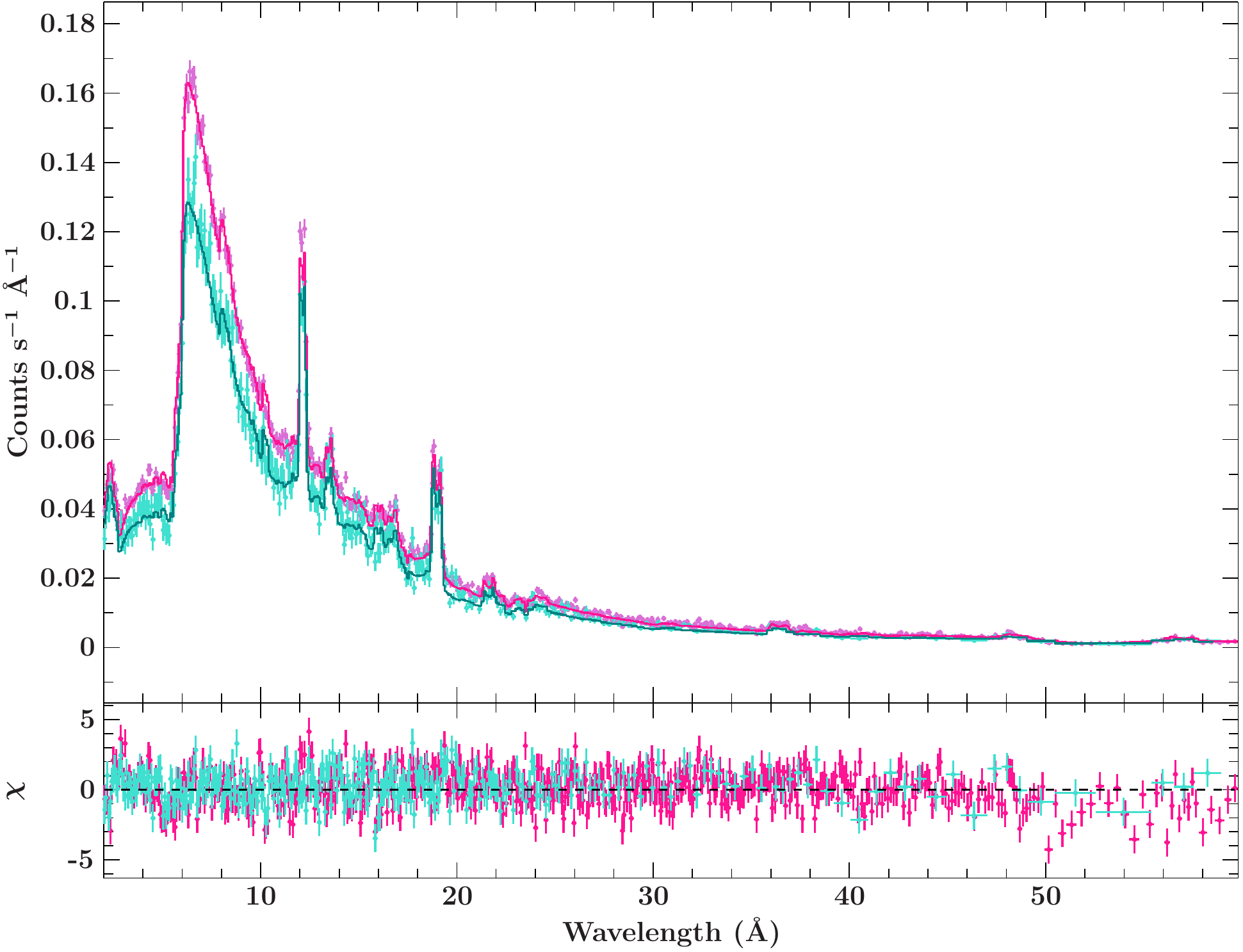}
  \caption{The LETGS spectra with the best-fitting model, consisting of a
    two-temperature APEC with increased neon abundance, blurred by an accretion
    disk. ObsID 16636 is lower-flux and plotted in turquoise, ObsIDs 15765 and 16637
    are brighter and plotted in pink. For clarity, we have rebinned the spectra
    and combined ObsIDs 15765 and 16637, as they are spectrally very similar.
    The apparent emission features at $\sim$36, $\sim$48, and $\sim$57\,\A are
    the \nex and \oviii lines in the higher grating orders.}
  \label{fig:bestfit}
\end{figure*}

\begin{figure*}
  \plotone{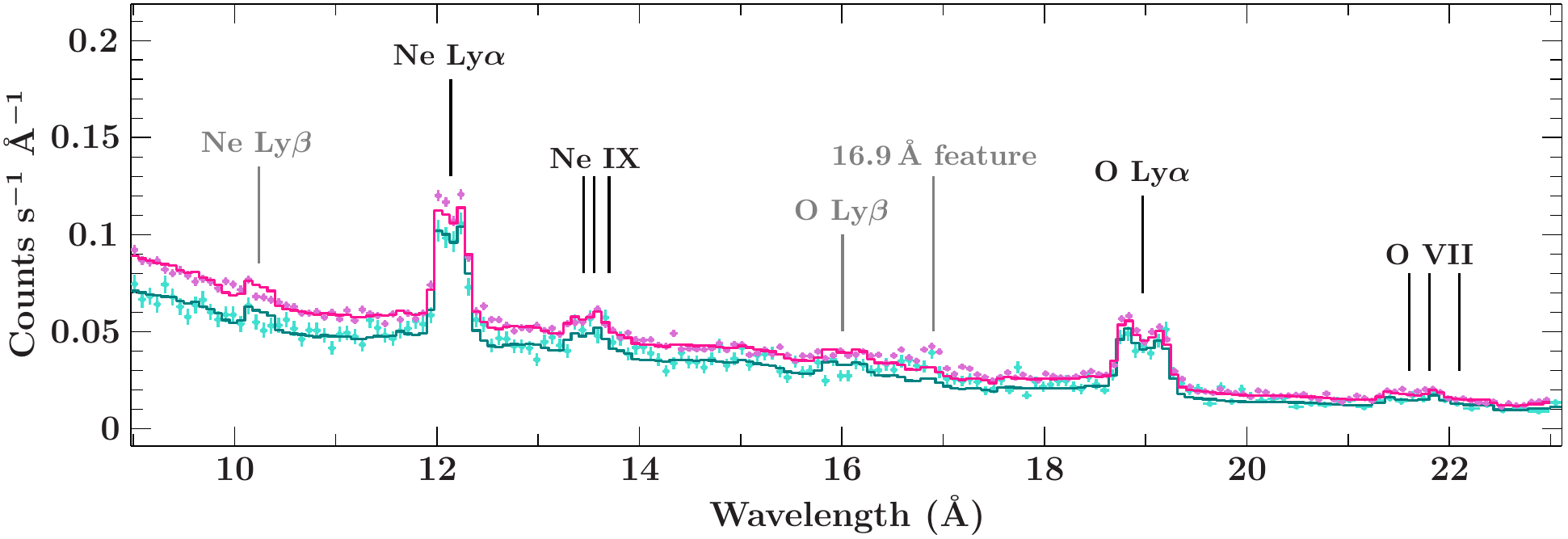}
  \plotone{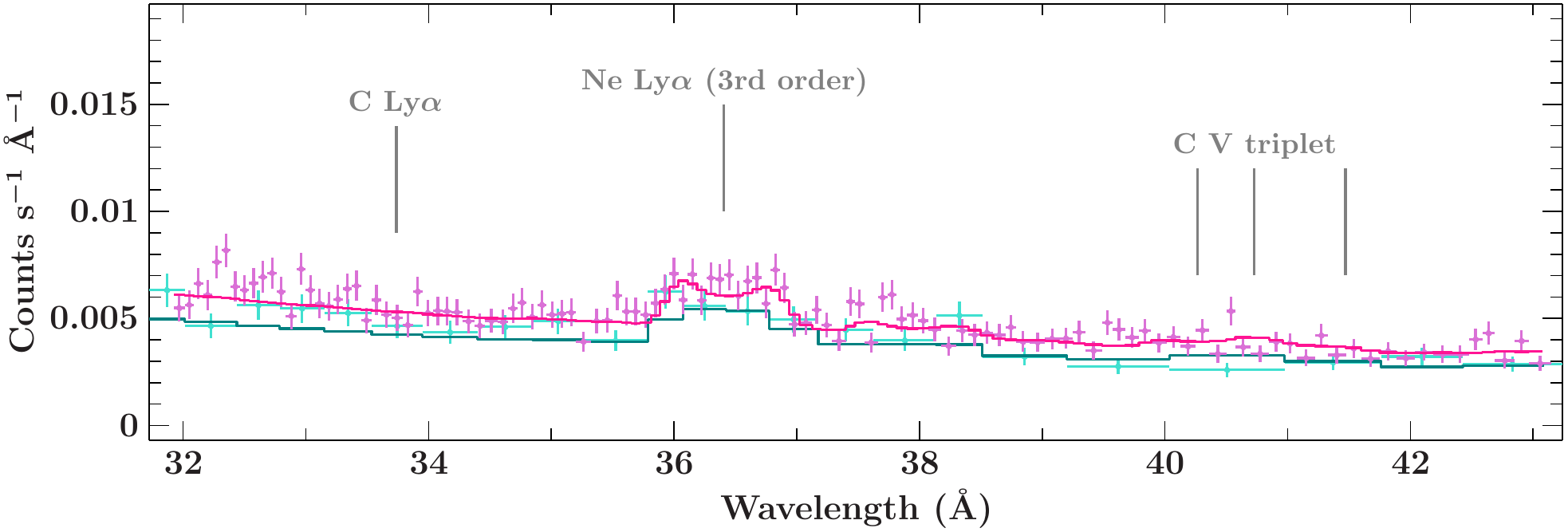}
  \caption{Emission line regions of interest, fitted with our best-fit
    two-temperature disk-blurred APEC model. Top: the neon and oxygen line
    regions. Bottom: the carbon region.  Both plots show ObsID 16636\ in
    turquoise and ObsIDs 15765 and 16637 combined in pink. \nex Ly$\beta$ is
    overproduced by the model and is not detected in the spectrum. The extra
    emission around $16.9$\,\A is of unknown origin. Carbon is not detected in
    emission in the X-ray spectrum, consistent with the high temperatures found
    by the APEC fit. The clear feature at $\sim$36.5\,\A is the \nex line in
    the third order; the second-order \oviii and third-order \neix lines are
  also visible in the model at $\sim$38 and 40--42\,\A.}
  \label{fig:line_regions}
\end{figure*}

The best-fitting model is dominated by a $\sim$13\,MK component,
which produces most of the neon emission and some of the \oviii feature, while
a $2.3$\,MK component mostly produces the helium-like oxygen lines. The
inclusion of the \texttt{rdblur} model simultaneously constrains the accretion
disk parameters by modeling the line profiles. These results can be found in Table
~\ref{tab:aped_rdblur}, and the spectrum with the best-fit model is plotted in
Figure~\ref{fig:bestfit}. Zoomed-in plots showing the line regions of interest
are plotted in Figure~\ref{fig:line_regions}. Compared to the unblurred APEC model of
Section~\ref{sec:aped}, these fits find considerably smaller turbulent
velocities, as blurring by the disk accounts for most of the line widths. The
disk parameters are largely consistent with what we found fitting the
individual lines with \texttt{diskline}.

In terms of chemical abundances, we find similar results to our pure APEC fits:
overabundant neon, poorly-constrained carbon, and a small but nonzero iron
abundance. The carbon and magnesium abundances are constrained to upper limits
of $0.488$ and $0.190$, although we repeat our previous caution that the carbon
abundance should be viewed as marginally reliable at these temperatures.
The most abundant iron species at 13 and 2.3\,MK are \ion{Fe}{23} and
\ion{Fe}{17}, respectively; the brightest iron lines in the APEC model, which
drive the nonzero abundance we observe, are the \ionrange{Fe}{17}{23} features,
with \ion{Fe}{21} at 12.284\,\A and \ion{Fe}{22} at 11.77\,\A flanking the \nex
line. As was found with the APEC fits in the previous section, the 17\,\A
feature is still present in the residuals and is not reproduced by the APEC
model; we were unable to find an APEC-based model which produced a solitary
17\,\A line, even with adding extra temperature components and allowing for
different abundances between the components.

\input{rdblur_CONeMgFe_allobs}

As with the continuum analysis, we also carried out a MCMC analysis of the
\texttt{rdblur}-APEC model, with 100 walkers for each free parameter and a
chain length of 2500. Most of the parameters are entirely consistent with the
best-fit model in Table~\ref{tab:aped_rdblur}. The only notable difference
between the MCMC results and the standard spectral fits is the carbon
abundance, where the MCMC runs converge to a high abundance of $\sim$2.2,
compared to the upper limit of $\sim$0.5 found in our spectral fits. This
reinforces our call for caution when interpreting the carbon abundances, due to
the high temperatures producing intrinsically weak carbon features.

\begin{figure}
  \centering
  \includegraphics[width=7.4cm]{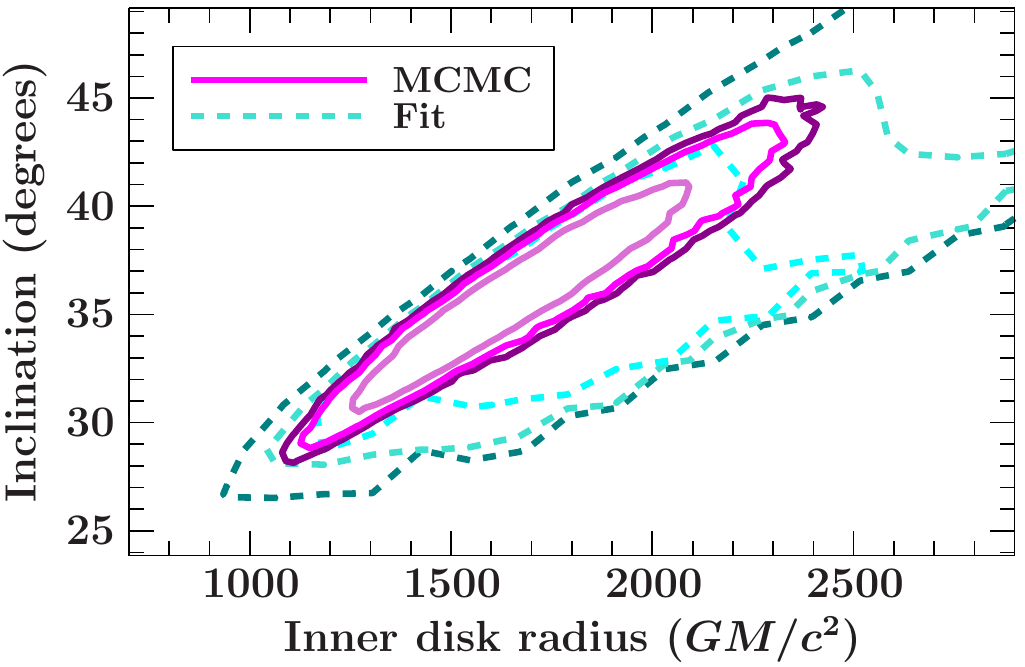}
  \caption{MCMC probability contours (magenta, solid) and
    traditionally-calculated confidence contours (blue, dashed) for inclination
    vs. inner disk radius for our \texttt{rdblur}-APEC model. From inside
    out, both show the 68\%, 90\%, and 95\% contours. The minimum inclination
    allowed by the 95\% contours is $\sim$25$^{\circ}$, in conflict with some of
    the very low inclinations previously suggested for the system.}
  \label{fig:incl_vs_rin}
\end{figure}

The posterior probability distributions resulting from the MCMC runs are useful
in studying correlations between the fit parameters. For instance, in
Figure~\ref{fig:incl_vs_rin}, we plot the posterior distribution for the inner
disk radius and the inclination. The strong correlation is not unexpected, as
the \texttt{diskline} and \texttt{rdblur} models are mostly sensitive to
$R_{\rm in}\sin i$. However, the contours cut off for inclinations less than
$\sim$27$^{\circ}$, ruling out lower inclinations for the source.
Figure~\ref{fig:incl_vs_rin} also shows confidence contours calculated by
freezing parameters and re-fitting; these are somewhat broader than the MCMC
results, but come to largely the same conclusion.

\section{Discussion} \label{sec:discussion}

\subsection{The accretion disk and system inclination} \label{sec:discuss_disk}

The double-peaked lines of neon and oxygen are easiest to explain as originating in
a disk, as other explanations typically require finely-tuned accretion
geometries. For example, a bidirectional outflow could produce a two-horned
line profile, but it should show a much dimmer red wing compared to the blue, as the
redshifted outflow would be obscured by the accretion disk or stream. Our
plasma modeling of these LETGS datasets provide excellent constraints on the
disk parameters, as it fits all lines and all spectral orders simultaneously.
The inner edge of the disk, if it is tracked by the emission lines, is at a
radius of $\sim$1500\,\gmc, or $3.1\times 10^{8}$\,cm for a 1.4\,\msol neutron
star. As noted above, this value is dependent on the inclination of the source,
but based on Figure~\ref{fig:incl_vs_rin}, it is unlikely to be smaller than
1000\,\gmc. The LETGS data also provide the best limits to date on the
outer radius of the emitting region, at $4300^{+3400}_{-1700}$\,\gmc; however,
it should be noted that this value is more sensitive to the choice of $q$,
which we fixed in this work.

In addition to the lower limits on the line profiles and inclination shown in
Figure~\ref{fig:incl_vs_rin}, we can also note that, given the pulsar's
established spin-up trend \citep{camero-arranz_new_2010}, the inner disk should
lie inside of the corotation radius, which at \six's 7.67\,s pulse period is
$6.5 \times 10^{8}$\,cm, or $\sim$3200\,\gmc. Extrapolating from the trend in
Figure~\ref{fig:incl_vs_rin}, this implies that the inclination cannot be
larger than 50--60$^{\circ}$. Thus, we can conservatively conclude that,
assuming the lines track the inner disk edge, the inclination likely lies in
the 25--60$^{\circ}$ range.

To investigate whether the emission lines really do track the inner edge of the
disk, we can check that they are consistent with where we expect the inner disk
to lie. The accretion disk is truncated by the neutron star's magnetic field at
the magnetospheric radius, $R_{\rm m}$, which theoretical arguments suggest
should lie within the \alfven radius, $R_{\rm A}$, given in
\citet{lamb_model_1973}, via \cite{becker_thermal_2007}, as:
\begin{align}
  R_{\rm A} =\ &2.29 \times 10^{8}
    \left( \frac{M_{\rm NS}}{1.4\,\msol} \right)^{-1/7}
    \left( \frac{R_{\rm NS}}{10{\rm\,km}} \right)^{12/7} \nonumber \\
    &\times \left( \frac{B}{10^{12}{\rm\,G}} \right)^{4/7} 
    \left( \frac{\dot{M}}{10^{17}{\rm\,\gs}} \right)^{-2/7} {\rm\,cm},
    \label{eqn:ra}
\end{align}
where $B$ is the magnetic field strength of the star (here assumed to be
dipolar), $\dot{M}$ is the accretion rate, and $M_{\rm NS}$ and $R_{\rm NS}$
are the mass and radius of the neutron star. Assuming an accretion rate of $1.1
\times 10^{16}$\,g\,s$^{-1}$ as found by \citet{schulz_collisional_2019}, a
dipole magnetic field strength of $4.2 \times 10^{12}$\,G, based on the CRSF
energy measured by \citet{dai_broadband_2017}, and a canonical 1.4\,\msol,
10\,km radius neutron star, $R_{\rm A}$ is approximately $10^{9}$\,cm, or
$\sim$4800\,\gmc. Our inner disk radius of $\sim$1500\,\gmc is thus
$\sim$0.3$R_{\rm A}$, in line with theoretical predictions and similar to
estimates of $R_{\rm m}/R_{\rm A}$ for other sources
\citep[e.g.,][]{filippova_radius_2017}.

\subsection{Plasma modeling}\label{sec:discuss_aped}
\subsubsection{Chemical abundances in \six}\label{sec:discuss_abund}

Our APEC fits give us a useful handle on the chemical abundances in the system.
However, the abundances reported in Table~\ref{tab:aped_rdblur} are relative to
the solar abundances of \citet{anders_abundances_1989}, while the donor star is
hydrogen-depleted. Thus, we measure the elemental abundances relative to oxygen
by fixing the oxygen abundance to unity. Following \citet{schulz_collisional_2019}, it
is straightforward to determine the number density ratio between some chosen
element X and oxygen:
\begin{equation}
  \frac{n_{\rm X}}{n_{\rm O}} = \frac{\rm Abund(X)}{\rm Abund(O)}\left(\frac{n_{\rm X}}{n_{\rm H}}\right)_{\rm AG89}\left(\frac{n_{\rm O}}{n_{\rm H}}\right)_{\rm AG89}^{-1},
  \label{eqn:abund}
\end{equation}
where $n_{\rm X}$ is the number density of element X, Abund(X) is the
abundance for the same element returned by APEC, and $\left(n_{\rm X}/n_{\rm
H}\right)_{\rm AG89}$ is taken from the \citet{anders_abundances_1989}
abundance tables.

\begin{figure*}
  \plotone{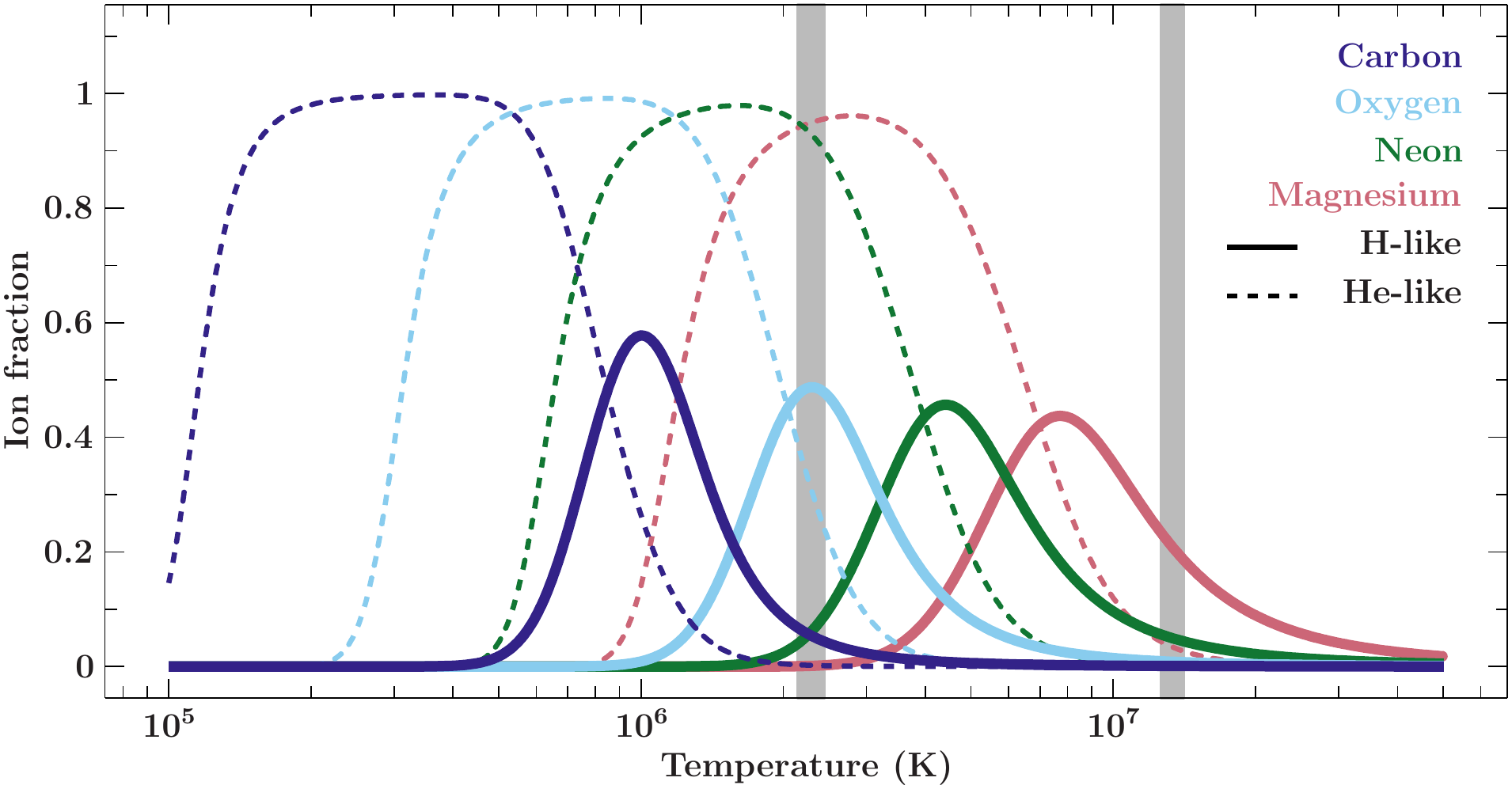}
  \caption{The ionization fractions of H-like and He-like carbon, oxygen,
  neon, and magnesium for a range of temperatures. The hydrogen-like ions are
  plotted with solid lines, and the helium-like ions are plotted with dashed
  lines. Our measured temperatures of $13$ and $2.3$\,MK are indicated by the
  gray shaded regions. Carbon is almost fully ionized at 2\,MK and completely
  ionized at 13\,MK, explaining its lack of presence in the X-ray spectrum.
  Magnesium, on the other hand, should have a significant fraction in the
  \ion{Mg}{12} state, meaning the lack of emission from magnesium is better
  explained by a lack of magnesium in the system.}
  \label{fig:ion_frac}
\end{figure*}

From the upper limits and measurements found in Table~\ref{tab:aped_rdblur}, we
find $n_{\rm Ne}/n_{\rm O} = 0.47 \pm 0.04$, $n_{\rm Mg}/n_{\rm O} \lesssim
0.0085$, and $n_{\rm Fe}/n_{\rm O} = 0.0042 \pm 0.0008$. For carbon, we find
$n_{\rm C}/n_{\rm O} \lesssim 0.21$ from our spectral fits and $n_{\rm
C}/n_{\rm O} \sim 0.93$ from our MCMC results. However, as noted earlier, this
discrepancy probably arises from the non-detection of any carbon features at
all. More quantitatively, Figure~\ref{fig:ion_frac} displays the fraction of
ions in the H-like and He-like states as a function of temperature, as
calculated by AtomDB. At our best-fit temperatures (the shaded regions in
Figure~\ref{fig:ion_frac}), carbon is mostly in the fully-ionized state,
producing intrinsically weak carbon features. In contrast, magnesium should be
mostly in the H- and He-like states, so the lack of Mg lines, under the
collisional plasma assumption, is better explained by an intrinsically low
abundance. We do not plot ionization fractions for iron, but as noted in
Section~\ref{sec:aped_rdblur}, at these temperatures iron is mostly in the
\ion{Fe}{17} and \ion{Fe}{23} states --- a range of ions with plenty of X-ray
lines in the LETGS' band.

Specifically on the iron abundance,
\citet{koliopanos_luminosity-dependent_2016} carried out an extensive study of
\six's Fe~K$\alpha$ fluorescence line, finding an inner disk radius from the
line width of 3.7--$15.7 \times 10^{8}$\,cm (1800--7600\,\gmc) --- somewhat
larger but not inconsistent with our measurements. However, they also derived a
considerably lower iron abundance, quoting an O/Fe ratio approximately 68 times
the solar value, while our measurement implies a ratio between 10 and 15 times
solar. There are likely significant systematic effects that both analyses
ignore --- our measurement stems from the lines hiding in the wings of the
\ion{Ne}{10} feature and upper limits on lines in the 15\,\A region in what is
assumed to be a fully collisionally-ionized, optically thin plasma (in reality,
the optically thin assumption may not hold and photoionization likely plays a
not insignificant role, \textit{viz.} the lack of a \nex K$\beta$ and the
17\,\A feature discussed in Section~\ref{sec:17A_line}), while
\citet{koliopanos_luminosity-dependent_2016} assume a C/O-dominated plasma, a
lamppost geometry, and a \citet{shakura_black_1973} disk \citep[the disk is
likely hotter hotter and heated by X-ray heating from the pulsar, clearly
contains significant neon, and may contain helium, per the discussion
in][]{schulz_collisional_2019}.

\subsubsection{The 16.9\,\A feature and collisional vs. photoionization}\label{sec:17A_line}

While our disk-blurred collisional plasma model fits these datasets quite well,
it fails in two main areas: first, the \nex K$\beta$ feature is overpredicted
by the APEC plasma model, and second, there is additional emission around
16.9\A that is consistently unmodeled. Furthermore, as found by
\citet{schulz_collisional_2019} and in Section~\ref{sec:aped}, the two
temperature components of the APEC plasma appear to originate around the same
radial distance, suggesting some sort of complex geometry (e.g., a clumpy
environment or an azimuthal temperature distribution) or that the
two-temperature model is only an approximation for a more complex system. These
give us some insight into the true nature of the plasma.

\paragraph{The \nex K$\beta$ line}
For this undermodeled feature at 10.24\,\A, the discrepancy between model and data
may lie in the APEC model's assumption of an optically-thin plasma. The \nex
K$\beta$ photons are more sensitive to optical depth --- in a more optically
thick plasma, they can be scattered into Ly$\alpha$ and UV photons, while
Ly$\alpha$ photons have no alternative transitions. Thus, careful simulations
of a plasma with a higher optical depth may obtain a better fit in this area.

\paragraph{The 16.9\,\A feature}
The excess emission at 16.9\,\A is somewhat more perplexing. It appears in all three
datasets and in both the positive and negative orders, and does not appear to
be a second- or third-order artifact (as there does not appear to be excess
emission at 8.45 or 5.63\,\A). Considering the instrument calibration, the
LETGS' effective area curve does have the two M edges of cesium to either side
of this feature, but the excess flux is quite high for a calibration feature.
Fitted with a Gaussian, the equivalent width of the feature is 12\,eV in the
first observation and 10\,eV in the last two, and the velocity width is
$\sim$5200\,\kms. However, it is also acceptably well-fit by a
\texttt{diskline} profile centered slightly red- or blueward of the peak. The
feature seen in the LETGS is too dim to be detected in the HETGS observations
investigated in \citet{schulz_collisional_2019}.

\begin{figure}
  \plotone{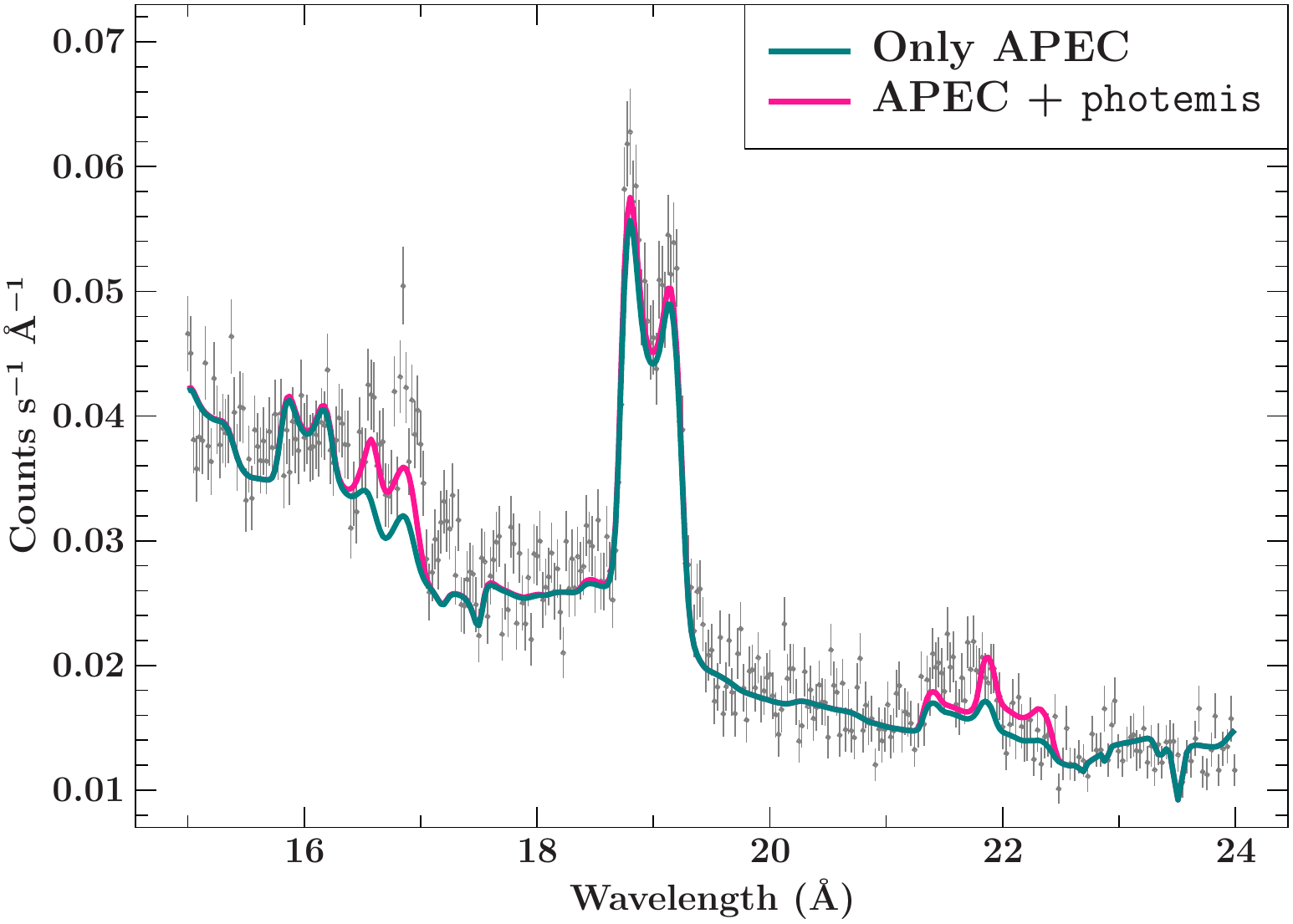}
  \caption{The oxygen region of the LETGS spectrum, comparing our two-temperature APEC model with and without an additional \texttt{photemis} component. The oxygen RRC, when blurred by the disk, begins to reproduce the excess emission at 16.9\,\A}
  \label{fig:apec_plus_photemis}
\end{figure}

There are no strong emission lines around 17\,A that would not be accompanied
by brighter lines at other wavelengths (e.g., \ion{Fe}{17} should also show
emission at 15 and 16\,A). Variations on our collisional plasma model (e.g.,
adding a third temperature component and/or more aggressive variation of
elemental abundances) fail to reproduce the extra emission. However, the
radiative recombination continuum (RRC) of oxygen does lie around 16--17\,A.
Our APEC-based model does not produce strong RRCs due to the high temperature,
but we find that adding a \texttt{photemis} component with $\log\xi \sim 1$ and
the same abundances as our APEC plasma produces a clear oxygen RRC which, when
blurred by the disk, recovers some of the excess emission around 16.9\,\A
\citep[while avoiding the strong neon RRC that appeared in the pure
\texttt{photemis} fits of][]{schulz_collisional_2019}. The comparison between
our pure APEC model and an APEC+\texttt{photemis} model is shown in
Figure~\ref{fig:apec_plus_photemis}. It should be noted that this is somewhat
of an ad-hoc approach to the problem and we do not claim that the
\texttt{photemis} component reflects the actual physical reality of the system
--- not only does \texttt{photemis} assume a far lower density
($10^{4}$\,cm$^{-3}$) and temperature (4800\,K) than we see in \six, but the
normalization of $\sim$100 required to produce the model displayed in
Figure~\ref{fig:apec_plus_photemis} implies an emitting region several orders
of magnitude larger than the entire \six system. Nonetheless, this should
provide something of a starting point for more self-consistent modeling of this
source, and at least demonstrates that an oxygen RRC, when blurred by the disk,
can reproduce the shape of the observed feature at 16.9\,\A. It is also
interesting to note that our ionization parameter, $\log\xi \sim 1$, is similar
to the ionization parameter predicted in \citet{schulz_collisional_2019}.

\subsection{On the nature of the donor star} \label{sec:discuss_donor}

\begin{figure}
  \plotone{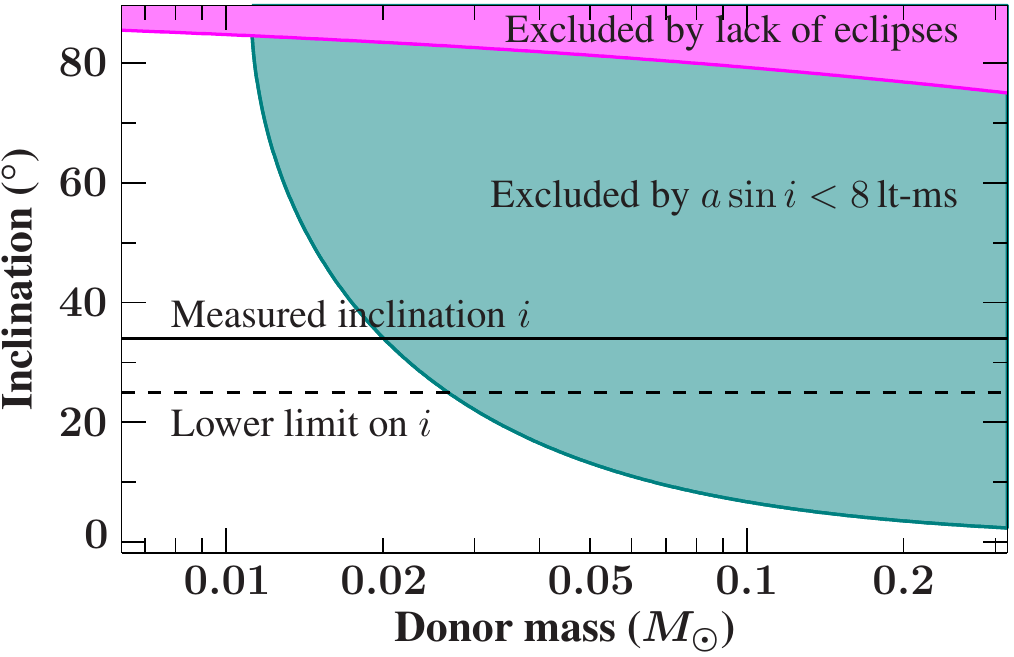}
  \caption{Limits on the inclination and donor mass in \six. The upper magenta
  shaded region indicates the upper limit on the inclination from the lack of
  eclipses in the system, based on the Roche lobe radius for the donor at that
  mass. The large teal shaded region is excluded due to the measured limit on
  the semimajor axis of $a\sin i < 8$\,lt-ms. Our measured inclination is
  indicated by a solid black line, and the approximate lower limit on the
  inclination from our confidence contours and MCMC runs is indicated by a dashed
  black line.}
  \label{fig:incl_vs_mass}
\end{figure}

In Figure~\ref{fig:incl_vs_mass}, we plot the relationship between inclination
and donor mass, assuming a 1.4\,\msol neutron star, based on the
observed lack of eclipses in the source and the published limit of $a\sin i <
8$\,lt-ms from \citet{shinoda_discovery_1990}. We assume a circular, 42-minute orbit
and that the donor fills its Roche lobe, the radius of which we take from
\citet{eggleton_aproximations_1983}. The relationship between inclination and
orbit size is from \citet{chakrabarty_high-speed_1998}. Our measurement of $i
\gtrsim 27^{\circ}$ thus implies a donor mass $M_{\rm d} \lesssim
0.026$\,\msol; higher inclinations would require progressively lower-mass
donors. This is independent of the precise nature of the donor --- as long as
the donor fills its Roche lobe and the disk inclination reflects the orbit
inclination, these constraints must be met. The question, then, is what type of
companion can reach this low of a mass, fill its Roche lobe, produce the
chemical abundances we observe, and provide a high enough accretion rate to
explain the spin-up and X-ray flux?

UCXBs, outside of globular clusters, are generally believed to follow one of
three possible formation channels, based on three different classes of donor
stars: hydrogen-depleted main-sequence stars, helium-burning stars, and white
dwarfs. As noted by \citet{nelemans_chemical_2010}, the donor in any UCXB is
almost certainly \emph{some} sort of degenerate, low-mass dwarf, as in the
main-sequence and helium-burning cases, the degenerate core of the star will be
exposed by mass transfer. Thus, the differences between these donors will be
mainly observable in their chemical composition and mass.

\citet{verbunt_evolutionary_1990} and \citet{chakrabarty_high-speed_1998}
rejected the helium-burning donor case, as this would either be orders of
magnitude brighter in the optical or orders of magnitude dimmer in the X-rays.
For the evolved main-sequence donor, while \six's low hydrogen fraction does
present a problem for this case, \citet{nelemans_chemical_2010} state that such
donors could have hydrogen fractions as low as $\sim$10$^{-5}$, and while the
hydrogen fraction in \six is certainly low \citep[$\lesssim$0.1, per
][]{werner_vlt_2006}, it is not clear that it is low enough to fully rule this
case out. However, per \citet{verbunt_evolutionary_1990}, the mass-radius
relationship for such an object means it would need to have a mass of about
0.08\,\msol to fill its Roche lobe at \six's orbital period, which is quite
strongly ruled out by our limits on the inclination. 

Thus, we are left with a white dwarf (that is, some kind of compact stellar
remnant supported primarily by electron degeneracy pressure) as the donor. As
noted by \citet{deloye_white_2003}, the possible types of white dwarf donors
exist more on a continuum than as a discrete set of classes, defined mainly
by the mass of the original main-sequence star and its age at the onset of mass
transfer --- a system that began mass transfer while the donor was burning
helium will have a different composition compared to one that began after a
white dwarf had formed.

Previous work has favored C/O or O/Ne compositions for the donor, based
on the bright neon and oxygen lines and the lack of detected helium. However, 
both C/O and O/Ne white dwarfs present issues from the composition standpoint. 
Taking the C/O case, with equal parts carbon and oxygen, the neon fraction in 
\six must be $\sim$0.2--0.3 based on our spectral fits. This is in fairly strong 
disagreement with models of C/O white dwarfs, which typically predict neon 
fractions of at most a few percent. Even if one considers chemical fractionation 
\citep[e.g.,][]{segretain_cooling_1994}, the neon fraction is a factor of a few 
too large. On the O/Ne side, the neon fraction is no longer an issue, but we find
the opposite problem with magnesium --- the magnesium fraction in an O/Ne white 
dwarf should be of order 10\%, while we constrain it to below a percent. 
Furthermore, in  \citet{schulz_collisional_2019}, we argued that helium must be 
present in order to keep the emitting volume within reasonable limits, given the 
plasma emission measure we obtain; however, this must be brought into agreement 
with the lack of detected helium in the visible and UV
\citep{homer_ultraviolet_2002,werner_vlt_2006} --- but it should also be noted 
that \citet{werner_vlt_2006}'s work failed to detect neon in the visible 
spectrum, in conflict with our clearly detected neon features. The work
presented here and in \citet{schulz_collisional_2019} provide very strong limits
on the mass of the donor star, the geometry of the accretion disk, and the 
chemical abundances in the system, but considerably more sophisticated plasma and
accretion disk modeling is needed to properly interpret these results.

\section{Conclusion}\label{sec:conclusion}

Our detailed analysis of the LETGS dataset on \six converges on a picture of an
accretion disk which is inclined at $\sim$34$^{\circ}$ (somewhat high relative
to previous estimates) and dominated by collisional ionization (although with
some contribution from photoionization), with a large overabundance of neon
relative to oxygen and underabundances of magnesium and iron. Our measurements
of the inner edge of the accretion disk constrain it to a lower limit of
$\sim$1000\,\gmc, or $2\times 10^8$\,cm, which is approximately 20\% of the
neutron star's \alfven radius and well within its corotation radius. Based on
the emission line profiles and \six's spin-up trend, the inclination is
constrained to between 25 and 60$^{\circ}$. The donor star must be very
low-mass, $\lesssim$0.026\,\msol, in order to meet the constraints on the size
of the system, and its chemical composition and plasma characteristics suggest
it is a helium white dwarf. In the following, we summarize some of the
remaining issues with this system and attempt to suggest some ways forward.

\subsection{Mass transfer rate}\label{sec:conclusion_mdot}

Evolutionary scenarios for UCXBs generally find systems reaching a minimum
period of $\sim$10\,minutes relatively quickly, followed by a longer-timescale
migration outwards towards longer orbital periods. Many of these scenarios
\citep[e.g., those of][]{yungelson_formation_2002,nelemans_chemical_2010}
predict mass-transfer via gravitational radiation in the
$10^{-12}$--$10^{-11}$\,\msolyr range, typically 1--2 orders of magnitude too
low given \six's X-ray flux. With the 3.5\,kpc distance found by
\citet{schulz_collisional_2019}, this discrepancy is ameliorated somewhat, but
not entirely --- the source is still approximately an order of magnitude
brighter than mass transfer via gravitational radiation would predict.

There are some possible mechanisms for driving more mass transfer.
\citet{heinke_galactic_2013} argued that \six had to be a persistent accretor,
rather than a long-duration transient, based on estimating the total accreted
mass and arguing that this was larger than the maximum mass of a helium or
carbon-oxygen disk. The source's continued X-ray emission to this
day means these arguments are still valid even considering the shorter distance
from \citet{schulz_collisional_2019}. \citeauthor{heinke_galactic_2013}
presented models starting from helium-burning stars which fit \six's accretion
rate and orbital period, although more work would need to be done to make sure
that such a donor does not burn too much helium, as this would conflict with
the plasma emission measure arguments in \citet{schulz_collisional_2019}. Their
suggested donor's mass on reaching a 42\,min orbit may also be slightly high
compared to our limits presented here, although they only consider a few
possible models and there may be room in the parameter space in this regard.
Another solution might be found in the irradiation of the donor ---
\citet{lu_effects_2017} find that X-ray irradiation of very low-mass white
dwarf donors in UCXBs can drive long-duration outbursts, lasting as long as
centuries for $\sim$0.01\,\msol donors. These suggest very different natures
for \six: under the model presented by \citet{heinke_galactic_2013}, \six is a
truly persistent source; under \citet{lu_effects_2017}'s argument, it is a very
long-duration transient.

\subsection{Accretion disk composition and donor star}\label{sec:conclusion_abund}

Under the assumption of a two-temperature, collisionally-ionized plasma, we
find a Ne/O ratio that is $3.24^{+0.29}_{-0.28}$ times solar, corresponding to
a value of $0.47 \pm 0.04$. This model largely reproduces the observed spectrum
with the exception of an overproduced \nex Ly$\beta$ line and a
line-like feature at 16.9\,\A, which we argue in Section~\ref{sec:17A_line}
likely stems from an unmodeled photoionized component of the spectrum. In
reality, as argued in \citet{schulz_collisional_2019}, the disk of \six likely
sits in an intermediate state where the assumptions of common plasma models
such as APEC and \texttt{photemis} fail to fully describe the system. In
particular, the very high density suggested in \citet{schulz_collisional_2019}
approaches or exceeds the limit where APEC and XSTAR models are valid.

% Note: luminosity of a SS73 disk at 1.1e16 g/s is about 5e33 erg/s
Our large Ne/O ratio and small Mg/O ratio are difficult to reconcile with most
He-depleted white dwarf models --- we find too much neon for a C/O white dwarf,
which will have a neon fraction of only a few percent, while a O/Ne white dwarf
should have a magnesium fraction of 5--10\% \citep[see,
e.g.,][]{dominguez_formation_1993,garcia-berro_evolution_1997,gil-pons_formation_2001},
far in excess of the magnesium abundance we find. This is consistent with our
findings in \citet{schulz_collisional_2019}, where we argued based on the plasma
emission measure that very little helium burning could have taken place in
\six, thus concluding that the donor is likely a helium white dwarf. This,
however, must be reconciled with the lack of helium features in the optical
spectra --- \citet{werner_vlt_2006} placed a limit on the helium fraction of
$\lesssim$10\%. Of course, \citet{werner_vlt_2006} also found little evidence
for neon in the system, while in the X-rays the presence of neon is
incontrovertible, so there are clearly significant discrepancies between the
optical and X-rays. The specific physics of the disk are an obvious place to
look here --- for example, the \citet{shakura_black_1973} disk used as a
baseline by \citet{werner_vlt_2006}'s model, at their assumed accretion
rate\footnote{Coincidentally, \citet{werner_vlt_2006} use nearly the same
accretion rate for their pre-torque reversal analysis as we found for the
post-torque reversal data, due to the larger distance assumed (by nearly all
studies of \six) at the time.}, has a total luminosity of a few $\times
10^{33}$\,erg\,s$^{-1}$. The APEC component in our best-fit model, in
comparison, has a luminosity of $\sim 5 \times 10^{34}$\,erg\,s$^{-1}$ --- an
order of magnitude higher. Helium and neon could thus be in a higher ionization
state than considered by \citet{werner_vlt_2006} --- although this, in turn,
would need to be reconciled with the fact that \citeauthor{werner_vlt_2006}
do see oxygen features, which would be unexpected if temperatures were high
enough to ionize neon to the point of non-detection.

\subsection{Accretion disk parameters and inclination}\label{sec:conclusion_incl}

The strong double-peaked emission lines in the LETGS spectrum provide some of
the best constraints to date on the parameters of the accretion disk, and are
largely consistent with those found using the HETGS
\citep{schulz_collisional_2019}. While the inner disk radius and inclination are
obviously degenerate with each other (Figure~\ref{fig:incl_vs_rin}), we still
obtain good lower limits on the inner disk radius ($\sim$1000\,\gmc) and
inclination ($\sim$27$^{\circ}$). On the high-inclination end, the neutron
star's spin-up trend means the inner disk cannot be farther out than the
corotation radius, $\sim$3200\,\gmc, which, extrapolating from
Figure~\ref{fig:incl_vs_rin}, limits the inclination to $\lesssim 60^{\circ}$.

These bounds, with a best-fit value of ${34^{\circ}}^{+8}_{-5}$, indicate a
somewhat high inclination compared to other estimates. Most recently, in
\citet{iwakiri_spectral_2019}, we carried out a full general relativistic
raytracing analysis of the hard X-ray pulse profiles of \six, which converged
on an inclination of $9.5^{\circ} \pm 0.2$. However, as stated in
\citet{iwakiri_spectral_2019}, the large number of free parameters and
systematic factors in such an analysis means the model presented is not unique,
and there is at least one other solution at an inclination of 27$^{\circ}$
(albeit with considerably more complex beam patterns). Thus, we think
while this disagreement is interesting, it is not necessarily something to be
particularly concerned about at this moment.

However, it is of interest to note that the inclinations measured here and in
\citet{iwakiri_spectral_2019} have slightly different definitions: our
measurement is of the inclination of the inner edge of the accretion disk, assuming Keplerian
rotation, while \citeauthor{iwakiri_spectral_2019}'s is the inclination of the
spin axis of the neutron star. If the spin axis and accretion disk were
misaligned, this would produce a systematic difference between the two
measurements. Such a scenario would likely drive a warp in the inner disk, which
would precess at some rate and could explain \six's torque reversals and
quasi-periodic flaring (see Figure~\ref{fig:lc}). This is not a new idea ---
\citet{wijers_warped_1999} suggested that a warp in the inner disk could have
flipped through 90$^{\circ}$ to produce \six's 1990 torque reversal, and
\citet{chakrabarty_millihertz_2001} interpreted a 1\,mHz QPO in the optical as
arising from a precessing warp in the disk. Precession of the disk would
produce variation in the inclination measured by the emission lines; the fact
that we do not observe any significant changes in the line profiles between these
LETGS observations taken within a few days of each other tells us that if
precession is occurring, it is probably not on a timescale of days. Shorter
timescales are likely too short for \chandra to probe effectively; however,
future missions with high spectral resolution and larger effective area (e.g.,
\textit{Lynx}) may be able to investigate this.

\subsection{Accretion-induced collapse} \label{sec:aic}

\six is fundamentally an odd system, with an apparently-young neutron star
(long pulse period, strong magnetic field) in orbit with an apparently-old
donor. The rough timescale for UCXB formation is $\sim 10^9$\,yr
\citep{vanhaaften_evolution_2012}, while magnetic field decay is estimated to
be a $10^8$\,yr process \citep[e.g.,][]{bhattacharya_decay_1992}, although this
depends strongly on the initial conditions and mechanisms considered. One
solution that has been proposed to reconcile these is that the neutron star
formed via accretion-induced collapse, which could sidestep some
timescale-related issues. \citet{yungelson_formation_2002} theorized that a
1.2\,\msol ONeMg white dwarf could undergo AIC and produce a 1.26\,\msol
neutron star in a \six-like system after about 350\,Myr. This is an attractive
scenario at first glace: with a large amount of accretion in this system
falling onto the original white dwarf and not the neutron star, magnetic burial
could be avoided, and the additional time could possibly allow for
crystallization to produce the enhanced neon abundance we observe.

However, there are at least two serious issues with the AIC scenario. First, an
AIC event would produce a $\sim$160\,Myr gap before accretion resumed onto the
neutron star, during which the neutron star would cool. This could actually
enhance the decay of the magnetic field due to ambipolar diffusion: per
\citet{goldreich_magnetic_1992}, the timescale for this process is $\sim
3$\,Gyr at a core temperature of $10^8$\,K, but this drops below 100\,Myr at a
temperature of a few $10^6$\,K. Building on this in a recent paper,
\citet{cruces_weak_2019} found that a neutron star with an initial core
temperature of $3 \times 10^8$\,K and a $10^{12}$\,G field could drop to $\sim
10^8$\,G in less than 100\,Myr. Thus, it is not at all clear that a post-AIC
binary would still retain a highly-magnetized neutron star once accretion
resumes. Additionally, it may be difficult to evolve an AIC event into a
\six-like system with a short orbital period and very low-mass companion ---
\citet{tauris_evolution_2013}'s study of accretion-induced collapse found that
for close-orbit, post-AIC systems, the donors retain masses between 0.65 and
0.85\,\msol, considerably larger than \six's 0.02\,\msol donor.

A detailed study of this subject is obviously beyond the scope of this paper,
but our results here may provide some useful context and constraints for future
studies.

\acknowledgements
This research has made use of data obtained from the Chandra
Data Archive and software (CIAO, TGcat, and ISIS) provided by 
the Chandra X-ray Center (CXC). We have also made extensive use
of the ISISscripts, a collection of user scripts for ISIS
provided by MIT and ECAP/Remeis Observatory
(\url{http://www.sternwarte.uni-erlangen.de/isis/}).

\facility{\chandra X-ray Observatory (CXO)}
\software{ISIS \citep{houck_isis:_2000}, HEASOFT, CIAO}

\bibliography{refs}

\end{document}

%% file: diskline_hlike_local.tex
\begin{deluxetable*}{lrrrr}
  \tabletypesize{\footnotesize}
  \tablewidth{0pt}
  \tablecaption{\texttt{diskline} fits to \ion{Ne}{10} and \ion{O}{8} lines\label{tab:diskline_hlike}}
  \tablehead{ & & \colhead{16636}&\colhead{15765}&\colhead{16637}}
  \startdata
  \multicolumn{2}{r}{Emissivity index $q$\tablenotemark{*}} & $-3.60$ & & \\ 
    & Inclination\tablenotemark{\dag} ($^{\circ}$) & $34^{+19}_{-10}$ & & \\ 
  \midrule
    \ion{Ne}{10} & Norm ($10^{-3}$ ph\,cm$^{-2}$\,s$^{-1}$) & $2.25\pm0.20$ & $2.68\pm0.14$ & $2.72^{+0.16}_{-0.15}$ \\ 
    & Wavelength (\AA) & $12.134^{+0.011}_{-0.008}$ & $12.135^{+0.009}_{-0.010}$ & $12.130^{+0.008}_{-0.010}$ \\ 
    & $R_{\rm in}$ ($10^{3}$ $GM/c^2$) & $1.5^{+3.7}_{-0.6}$ & $1.4^{+3.4}_{-0.6}$ & $1.1^{+3.4}_{-0.5}$ \\ 
    % Line below is original, next has values pegged at upper bound rewritten
    % as lower limits
    % & $R_{\rm out}$ ($10^{3}$ $GM/c^2$) & $6^{+995}_{-4}$ & $10^{+25}_{-6}$ & $12^{+988}_{-9}$ \\ 
    & $R_{\rm out}$ ($10^{3}$ $GM/c^2$) & $> 2.4$ & $10^{+25}_{-6}$ & $> 3.5$ \\ 
  \midrule
    \ion{O}{8} & Norm ($10^{-3}$ ph\,cm$^{-2}$\,s$^{-1}$) & $2.59\pm0.30$ & $3.08\pm0.21$ & $3.00\pm0.22$ \\ 
    & Wavelength (\AA) & $18.977^{+0.016}_{-0.017}$ & $18.980\pm0.010$ & $18.965^{+0.014}_{-0.011}$ \\ 
    & $R_{\rm in}$ ($10^{3}$ $GM/c^2$) & $1.4^{+1.0}_{-0.5}$ & $1.4^{+3.3}_{-0.5}$ & $1.4^{+3.4}_{-0.7}$ \\ 
    & $R_{\rm out}$ ($10^{3}$ $GM/c^2$) & $3.3^{+2.5}_{-1.3}$ & $4.8^{+7.9}_{-2.2}$ & $7^{+21}_{-4}$ \\ 
  \midrule
    & $\chi^2_{\nu}$ (dof) & 1.22 (443) &  &  \\ 
  \enddata
  \tablenotetext{*}{$q$ frozen in fits}
  \tablenotetext{\dag}{Inclination tied between all ObsIDs}
\end{deluxetable*}

%% file: diskline_helike_local.tex
% \begin{deluxetable}{rrrrr}
%   \tabletypesize{\footnotesize}
%   \tablewidth{0pt}
%   \tablecaption{Line normalizations\tablenotemark{*} of \neix and \ovii triplets\label{tab:diskline_helike}}
%   \tablehead{& \colhead{$\lambda$ (\A)\tablenotemark{\dag}}&\colhead{16636}&\colhead{15765}&\colhead{16637}}
%   \startdata
%   \ion{Ne}{9} $r$ & 13.447 & $0.29\pm0.21$ & $0.22\pm0.16$ & $0.46^{+0.15}_{-0.19}$ \\ 
%     $i$ & 13.5515 & $0.30\pm0.25$ & $0.21\pm0.18$ & $< 0.276$ \\ 
%     $f$ & 13.699 & $0.25\pm0.19$ & $0.29\pm0.13$ & $0.19^{+0.14}_{-0.16}$ \\ 
%     $G$-ratio &   & $1.9\pm1.7$ & $2.3\pm1.9$ & $< 0.988$ \\ 
%     \midrule
%     \ion{O}{7} $r$ & 21.602 & $0.7\pm0.5$ & $1.3\pm0.4$ & $1.0\pm0.4$ \\ 
%     $i$ & 21.8025 & $0.5\pm0.5$ & $< 0.767$ & $< 0.798$ \\ 
%     $f$ & 22.098 & $< 0.411$ & $< 0.486$ & $< 0.565$ \\ 
%     $G$-ratio &   & $< 1.54$ & $< 0.805$ & $< 1.17$ \\ 
%     \midrule
%     $\chi^2_{\nu}$ (dof) &   & 1.17 (996) &  &  \\ 
%   \enddata
%   \tablenotetext{*}{Units: $10^{-3}$\,\pflx}
%   \tablenotetext{\dag}{Line wavelength, from \citet{drake_theoretical_1988}}
% \end{deluxetable}

\begin{deluxetable}{rrrrr}
  \tabletypesize{\footnotesize}
  \tablewidth{0pt}
  \tablecaption{Line normalizations\tablenotemark{*} of \neix and \ovii triplets\label{tab:diskline_helike}}
  \tablehead{& \colhead{$\lambda$ (\A)\tablenotemark{\dag}}&\colhead{16636}&\colhead{15765}&\colhead{16637}}
  \startdata
  \neix $r$ & 13.447 & $0.29\pm0.21$ & $0.22\pm0.16$ & $0.46^{+0.15}_{-0.19}$ \\ 
    $i$ & 13.5515    & $0.30\pm0.25$ & $0.21\pm0.18$ & $< 0.276$ \\ 
    $f$ & 13.699     & $0.25\pm0.19$ & $0.29\pm0.13$ & $0.19^{+0.14}_{-0.16}$ \\ 
    $R$\tablenotemark{\ddag} & $f/i$  & $0.5^{+3.4}_{-0.4}$ & $0.9^{+4.3}_{-0.5}$ & $0.5^{+6.0}_{-7.0}$ \\
    $G$ & $(f+i)/r$                   & $1.3^{+4.5}_{-0.7}$ & $1.8^{+4.5}_{-0.9}$ & $0.4^{+1.0}_{-0.3}$ \\
    \midrule
  \ovii $r$ & 21.602 & $0.7\pm0.5$ & $1.3\pm0.4$ & $1.0\pm0.4$ \\ 
    $i$ & 21.8025    & $0.5\pm0.5$ & $< 0.767$ & $< 0.798$ \\ 
    $f$ & 22.098     & $< 0.411$ & $< 0.486$ & $< 0.565$ \\ 
    $R$ & $f/i$      & $-0.1^{+1.8}_{-1.2}$ & $0.0^{+3.3}_{-1.5}$ & $0.1^{+3.9}_{-2.6}$ \\
    $G$ & $(f+i)/r$  & $0.4^{+2.4}_{-0.4}$  & $0.4^{+0.6}_{-0.2}$ & $0.5^{+1.0}_{-0.3}$ \\
    $\chi^2_{\nu}$ (dof) &   & 1.17 (996) &  &  \\ 
  \enddata
  \tablenotetext{*}{Units: $10^{-3}$\,\pflx}
  \tablenotetext{\dag}{Line wavelength \citep[fixed, from][]{drake_theoretical_1988}}
  \tablenotetext{\ddag}{90\% confidence intervals for $R$ and $G$ estimated from MCMC}
\end{deluxetable}

% From MCMC with lines limited to positive flux
% Ne IX
% $R$\tablenotemark{\ddag} & $f/i$      & $0.5^{+3.0}_{-0.3}$ & $1.0^{+3.8}_{-0.5}$ & $0.5^{+6.2}_{-0.3}$ \\
% $G$ & $(f+i)/r$                       & $1.4^{+4.2}_{-0.6}$ & $1.8^{+4.7}_{-0.7}$ & $0.7^{+1.6}_{-0.3}$ \\
% O VII
% $R$ & $f/i$      & $< 3.3$ & $0.22^{+3.28}_{-0.14}$ & $0.3^{+3.8}_{-0.2}$ \\
% $G$ & $(f+i)/r$  & $0.8^{+3.2}_{-0.4}$ & $0.5^{+0.7}_{-0.3}$ & $0.6^{+1.2}_{-0.3}$ \\

% From error propagation
% Ne IX
% $R$ & $f/i$      & $< 1.78$ & $1.4\pm1.3$ & $< 15.0$ \\ 
% $G$ & $(f+i)/r$  & $1.9\pm1.7$ & $2.3\pm1.9$ & $< 0.99$ \\ 
% O VII
% $R$ & $f/i$      & $< 0.414$ & $< 1.18$ & $< 1.57$ \\ 
% $G$ & $(f+i)/r$  & $< 1.54$ & $< 0.805$ & $< 1.17$ \\ 

%% file: rdblur_CONeMgFe_allobs.tex
\begin{deluxetable}{llr}
  \tabletypesize{\footnotesize}
  \tablewidth{0pt}
  \tablecaption{Two-temperature disk-blurred APED fit to \six\label{tab:aped_rdblur}}
  \tablehead{\colhead{}&\colhead{Parameter}&\colhead{Value}}
  \startdata
    Plasma & Norm$_{\rm hot}$ ($\times 10^{-2}$\,cm$^{-5}$)\tablenotemark{*} & $4.4\pm0.5$ \\ 
     & Norm$_{\rm cool}$ ($\times 10^{-2}$\,cm$^{-5}$) & $0.26\pm0.04$ \\ 
     & $T_{\rm hot}$ (MK) & $13.4^{+0.7}_{-0.9}$ \\ 
     & $T_{\rm cool}$ (MK) & $2.30^{+0.15}_{-0.16}$ \\ 
     & $v_{\rm turb, hot}$ (\kms) & $970^{+300}_{-390}$ \\ 
     & $v_{\rm turb, cool}$ (\kms) & $< 1210$ \\ 
     & C abundance\tablenotemark{\dag} & $< 0.488$ \\ 
     & O abundance & $1$ (fixed) \\ 
     & Ne abundance & $3.24^{+0.29}_{-0.28}$ \\ 
     & Mg abundance & $< 0.190$ \\ 
     & Fe abundance & $0.077\pm0.014$ \\ 
    \midrule
    Disk & $R_{\rm in}$ ($10^{3}$\,\gmc) & $1.5^{+0.6}_{-0.4}$ \\ 
     & $R_{\rm out}$ ($10^{3}$\,\gmc) & $4.3^{+3.4}_{-1.7}$ \\ 
     & Inclination ($^{\circ}$) & $34^{+8}_{-5}$ \\ 
    \midrule
    Continuum & \nh ($10^{22}$\,cm$^{-2}$) & $0.153\pm0.012$ \\ 
    16636 & $F_{0.5\text{--}10{\rm keV}}$ ($\times 10^{-10}$\,\eflx) & $4.63$ \\ 
    & Norm$_{\rm PL}$ ($10^{-2}$\,\pflx) & $3.4\pm0.4$ \\ 
     & $\Gamma$ & $1.16\pm0.07$ \\ 
     & Norm$_{\rm BB}$ ($R_{\rm km}^{2}/D_{\rm 10}^{2}$) & $101^{+29}_{-25}$ \\ 
     & $kT_{\rm BB}$ (keV) & $0.499^{+0.033}_{-0.028}$ \\ 
    15765 & $F_{0.5\text{--}10{\rm keV}}$ ($\times 10^{-10}$\,\eflx) & $5.47$ \\ 
    & Norm$_{\rm PL}$ ($10^{-2}$\,\pflx) & $4.0\pm0.4$ \\ 
     & $\Gamma$ & $1.19\pm0.06$ \\ 
     & Norm$_{\rm BB}$ ($R_{\rm km}^{2}/D_{\rm 10}^{2}$) & $128^{+19}_{-18}$ \\ 
     & $kT_{\rm BB}$ (keV) & $0.520^{+0.017}_{-0.014}$ \\ 
    16637 & $F_{0.5\text{--}10{\rm keV}}$ ($\times 10^{-10}$\,\eflx) & $5.52$ \\ 
    & Norm$_{\rm PL}$ ($10^{-2}$\,\pflx) & $4.1\pm0.4$ \\ 
     & $\Gamma$ & $1.19\pm0.06$ \\ 
     & Norm$_{\rm BB}$ ($R_{\rm km}^{2}/D_{\rm 10}^{2}$) & $119^{+21}_{-19}$ \\ 
     & $kT_{\rm BB}$ (keV) & $0.519^{+0.020}_{-0.018}$ \\ 
    \midrule
    & $A_{\rm G}$ ($10^{-4}$\,\pflx) & $8.7^{+1.4}_{-1.3}$ \\ 
     & $\lambda_{\rm G}$ (\AA) & $16.90^{+0.06}_{-0.05}$ \\ 
     & $\sigma_{\rm G}$ (\AA) & $0.29\pm0.05$ \\ 
    \midrule
     & $\chi^2_{\nu}$ (dof) & 1.11 (3883) \\ 
  \enddata
  \tablenotetext{*}{APEC normalization}
  \tablenotetext{\dag}{All abundances are relative to oxygen, as returned by APEC and assuming solar abundances; see Section~\ref{sec:discuss_aped} for interpretation.}
\end{deluxetable}